%% file: main.tex
\documentclass[sigconf, nonacm]{acmart}
\DeclareMathDelimiter{(}{\mathopen} {operators}{"28}{largesymbols}{"00}
\DeclareMathDelimiter{)}{\mathclose}{operators}{"29}{largesymbols}{"01}
\usepackage[a-1b]{pdfx}
\usepackage{xcolor}
\usepackage{amsmath}
\usepackage{makecell}
\usepackage{scalerel}
\usepackage{balance}
\usepackage{epstopdf}
\usepackage{graphicx}
\usepackage{hyperref} 
\usepackage{subfigure}
\usepackage{balance}
\usepackage{color}
\usepackage{tabularx}
\PassOptionsToPackage{hyphens}{url}\usepackage{hyperref}
\usepackage{ragged2e}  
\newcolumntype{Y}{>{\RaggedRight\arraybackslash}X}
\usepackage{booktabs}
\usepackage{algorithm}

\usepackage[noend]{algpseudocode}
\usepackage{fixltx2e}
\usepackage{listings}
\usepackage{multirow}
\usepackage{caption} 
\usepackage{comment}
\usepackage[labelfont=bf]{caption}
\MakeRobust{\Call}
\usepackage{enumitem} \setlist{nolistsep}
\usepackage{etoolbox} 
\usepackage{mdwlist} 

\usepackage{tcolorbox} 

\lstdefinestyle{sharpc}{language=[Sharp]C, frame=lr, rulecolor=\color{blue!80!black}}

\usepackage{color}
\definecolor{bluekeywords}{rgb}{0.13,0.13,1}
\definecolor{greencomments}{rgb}{0,0.5,0}
\definecolor{redstrings}{rgb}{0.9,0,0}

\usepackage{listings}
\newcommand{\val}[1]{\texttt{\small #1}}

\newcommand{\sj}{\textsc{Auto-Pipeline}}
\newcommand{\sjsl}{\textsc{Auto-Pipeline-SL}}
\newcommand{\sjrl}{\textsc{Auto-Pipeline-RL}}
\newcommand{\sjs}{\textsc{Auto-Pipeline-Search}}

\algnewcommand\algorithmicforeach{\textbf{for each}}
\algdef{S}[FOR]{ForEach}[1]{\algorithmicforeach\ #1\ \algorithmicdo}
\algnewcommand\Input{\item[\algorithmicinput]}%
\algnewcommand\Output{\item[\algorithmicoutput]}%
\algrenewcommand\textproc{}

\DeclareMathOperator*{\avg}{avg}
\DeclareMathOperator*{\argmax}{arg\,max}

\newcounter{definition}
\newenvironment{definition}[1][]{\refstepcounter{definition}\par\smallskip\textsc{Definition~\thedefinition.\ #1}}{\smallskip}

\newcounter{example}
\newenvironment{example}[1][]{\refstepcounter{example}\par\smallskip\textsc{Example~\theexample.\ #1}}{\smallskip}

\newcounter{theorem}

\newcounter{proposition}

\newtoggle{fullversion}
\toggletrue{fullversion}

\makeatletter
  \newcommand\figcaption{\def\@captype{figure}\caption}
  \newcommand\tabcaption{\def\@captype{table}\caption}
\makeatother

\newcommand\vldbdoi{10.14778/3476249.3476303}
\newcommand\vldbpages{2563 - 2575}
\newcommand\vldbvolume{14}
\newcommand\vldbissue{11}
\newcommand\vldbyear{2021}
\newcommand\vldbauthors{\authors}
\newcommand\vldbtitle{\shorttitle} 
\newcommand\vldbavailabilityurl{http://vldb.org/pvldb/format_vol14.html}
\newcommand\vldbpagestyle{empty} 

\begin{document}
\title{Auto-Pipeline: Synthesizing Complex Data Pipelines By-Target Using Reinforcement Learning and Search}

\author{Junwen Yang}
\affiliation{%
  \institution{University of Chicago}
}
\email{junwen@uchicago.edu}

\author{Yeye He}
\affiliation{%
  \institution{Microsoft Research}
}
\email{yeyehe@microsoft.com}

\author{Surajit Chaudhuri}
\affiliation{%
  \institution{Microsoft Research}
}
\email{surajitc@microsoft.com}
\begin{abstract}

       Recent work has made significant progress in helping users to 
       automate \textit{single} data preparation steps, such 
       as string-transformations 
       and table-manipulation operators 
       (e.g., Join, GroupBy, Pivot, etc.). We in 
       this work propose to automate \textit{multiple} 
       such steps end-to-end,
       by synthesizing complex data-pipelines
       with both string-transformations and 
       table-manipulation operators. 
       
       We propose a novel \textit{by-target} paradigm
       that allows users to easily specify the desired pipeline, 
       which is a significant departure from the 
       traditional \textit{by-example} paradigm. 
       Using by-target, users would provide
        input tables (e.g., csv or json files), and point us to
       a ``target table'' (e.g., an existing database table
       or BI dashboard) to demonstrate how the output 
       from the desired pipeline would schematically ``look like''. 
       While the problem is seemingly 
       under-specified, our unique insight is that 
       implicit table constraints such as FDs and keys can be exploited
       to significantly constrain the space and
       make the problem tractable. We develop an \sj{} system that 
       learns to synthesize pipelines using deep reinforcement-learning (DRL) and search.
       Experiments using a
       benchmark of 700 real pipelines crawled from GitHub and commercial vendors
			suggest that \sj{} can successfully synthesize
       around 70\% of complex pipelines with up 
       to 10 steps.\end{abstract}

\maketitle

\pagestyle{\vldbpagestyle}
\begingroup\small\noindent\raggedright\textbf{PVLDB Reference Format:}\\
\vldbauthors. \vldbtitle. PVLDB, \vldbvolume(\vldbissue): \vldbpages, \vldbyear.\\
\href{https://doi.org/\vldbdoi}{doi:\vldbdoi}
\endgroup
\begingroup
\renewcommand\thefootnote{}\footnote{\noindent
This work is licensed under the Creative Commons BY-NC-ND 4.0 International License. Visit \url{https://creativecommons.org/licenses/by-nc-nd/4.0/} to view a copy of this license. For any use beyond those covered by this license, obtain permission by emailing \href{mailto:info@vldb.org}{info@vldb.org}. Copyright is held by the owner/author(s). Publication rights licensed to the VLDB Endowment. \\
\raggedright Proceedings of the VLDB Endowment, Vol. \vldbvolume, No. \vldbissue\ %
ISSN 2150-8097. \\
\href{https://doi.org/\vldbdoi}{doi:\vldbdoi} \\
}\addtocounter{footnote}{-1}\endgroup



\input{introduction}

\input{problem}

\input{algorithm-search}

\input{algorithm-learning}

\input{experiment}
\input{related}

\input{conclusion}
\clearpage


\bibliographystyle{ACM-Reference-Format}
\balance
\bibliography{auto-pipeline}

\end{document}

%% file: introduction.tex
\section{Introduction}
\label{sec:intro}

\textit{Data preparation}, sometimes also known as data wrangling,
refers to the process of building sequences of table-manipulation
steps (e.g., Transform, Join, Pivot, etc.), to
bring raw data into a form that is ready for downstream applications (e.g., BI
or ML). The end-result of
data preparation is often a \textit{workflow} or \textit{data-pipeline}
with a sequence of these steps,
which are often then operationalized as recurring jobs in production.

It has been widely reported that business
analysts and data scientists spend 
a significant fraction of their time on data
preparation tasks (some report 
numbers as high as 80\%~\cite{Dasu03, deng2017data}). 
Accordingly, Gartner calls
data preparation ``the most time-consuming step
in analytics''~\cite{Sallam16}.
This is particularly challenging
for less-technical users, who increasingly need to
prepare data themselves today.


In response, significant progress has been made 
in the research community toward helping users
author \textit{individual} data preparation steps 
in data-pipelines. Notable efforts include
automated data transformations
(e.g.,~\cite{Abedjan16, gulwani2011automating, 
He2018TDE, heer2015predictive}), table-joins 
(e.g.,~\cite{search-join, auto-join}),
and table-restructuring (e.g.,~\cite{barowy2015flashrelate, Jin17, auto-suggest}), etc. 

In commercial systems, while pipelines are traditionally
built manually (e.g., using
drag-and-drop tools to build ETL pipelines),
leading vendors have adopted recent advances in research
and released features that 
make it really easy for users to build key steps in pipelines
(e.g., automated 
transformation-by-example has been used in Excel~\cite{TBE-Excel},
Power Query~\cite{TBE-pq}, and Trifacta~\cite{TBE-trifacta}; 
automated join has been used in
Tableau~\cite{AutoJoin-Tableau} and Trifacta~\cite{AutoJoin-Trifacta}, etc.).

\textbf{Automating multi-step pipeline-building.}
While assisting users to build \textit{single} data-prep steps
(e.g., Transform, Join, etc.) is great progress,
not much attention has been given to the more
ambitious goal of automating \textit{multi-step}
pipeline-building end-to-end. We argue that
building on top of recent success in automating single-steps
such as~\cite{auto-suggest}, synthesizing \textit{multi-step}
pipelines has become feasible and will be an area that warrants more attention.

The key challenge in multi-step pipeline-synthesis
is to allow users to easily specify the desired pipelines.
Existing methods use the ``by-example'' paradigm
(e.g., SQL-by-example~\cite{sql-by-example} and 
Query-by-output~\cite{qbo}), which unfortunately requires
a \textit{matching} pair of input/output tables to be
provided in order for the desired program (e.g., in SQL)
to be synthesized. While by-example
is easy-to-use for \textit{row-to-row}
string transformation~\cite{gulwani2011automating, 
He2018TDE} (because
users only need to type 2-3 example values), 
for \textit{table-to-table} transformations this paradigm
would unfortunately require users to manually enter an entire 
output table, which is not only significant overhead,
but can also be infeasible for users to provide in many cases
(e.g., when complex aggregations are required on large tables).

Furthermore, existing by-example approaches
largely resort to some forms of exhaustive search, which 
unfortunately limits the richness of the 
operators they can support, also making
these approaches frequently fail or time-out when synthesizing 
real pipelines with large amounts of data.

\textbf{New paradigm: ``by-target'' pipeline-synthesis.}
In this work, we propose a new paradigm for multi-step
pipeline-synthesis called \textit{by-target}. We show that a ``target''
is easy for users to provide, yet it still provides a sufficient specification
for desired pipelines to be synthesized. We emphasize that this
novel paradigm is not studied before, and is
a significant departure from the by-example approach.

Our key observation here is
that a common usage pattern in pipeline-building (e.g., ETL)
is to onboard new data files,
such as sales data from a new store/region/time-period, etc.,
that are often formatted differently. 
In such scenarios, users typically have a 
precise ``target'' in mind, such as an existing data-warehouse table,
where the goal is to bring the new data into a form 
that ``looks like'' the existing target table (so that the new data
can be integrated).
Similarly, in building visualizations and dashboards
for data analytics (e.g., in Tableau or Power BI), users can
be inspired by an existing visualization,
and want to turn their raw data into a visualization that ``looks like'' 
the given target visualization (in which case we can target
the underlying table of the visualization).

\begin{figure}[t]
\vspace{-12mm}
    \centering
    \includegraphics[width=1\columnwidth]{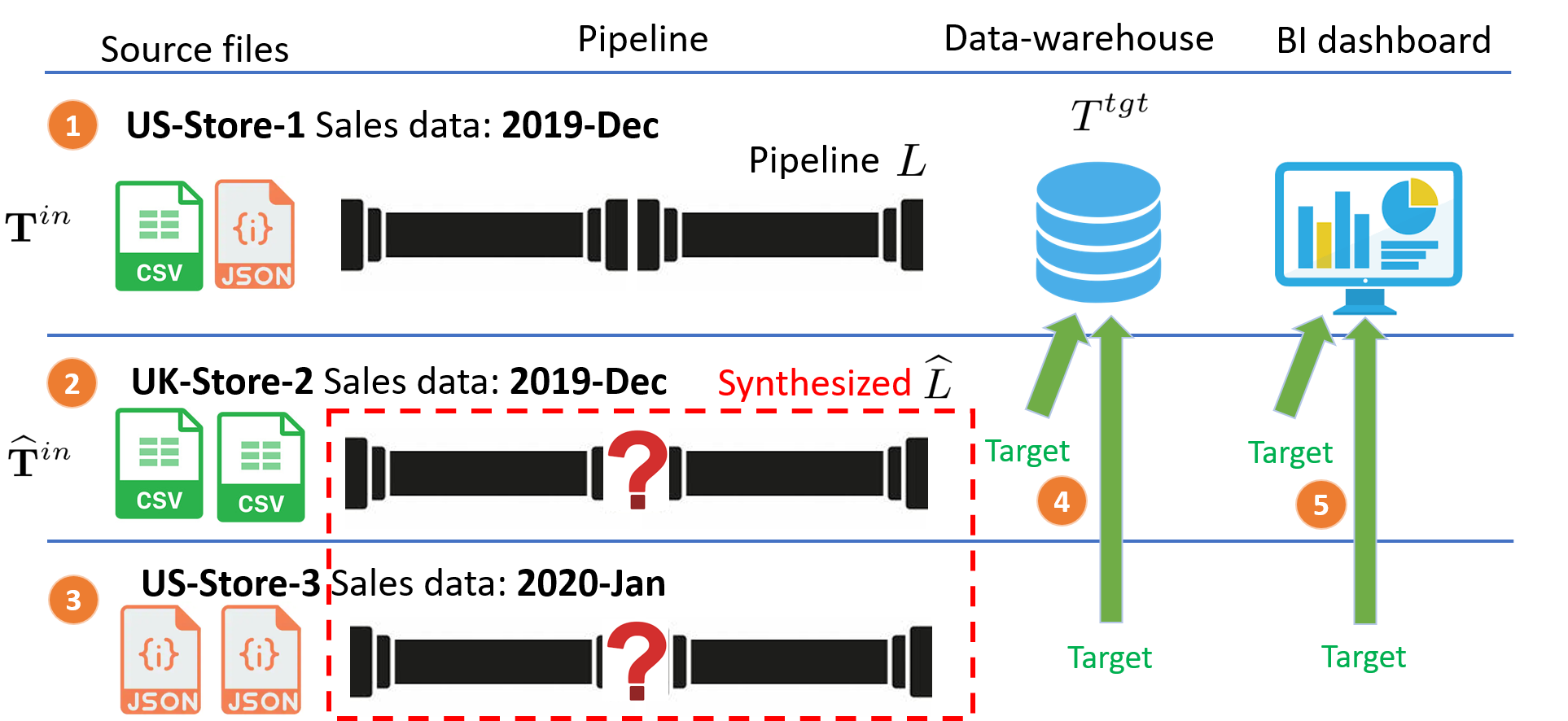}
    \vspace{-4mm}
    \caption{An example of pipeline-by-target. (1-3): Input tables
from different time-periods/store-locations often 
have different formats and schema. (1):
A pipeline previously built on one chunk of 
the input to produce database tables or BI dashboards. 
(2, 3): Instead of manually building pipelines
for new chunks of input, we try to synthesize these
pipelines by asking users to point us to a fuzzy ``target'' that can be
(4) an existing table or (5) an existing visualization.}
\vspace{-7mm}
\label{fig:by-target}
\end{figure}

Figure~\ref{fig:by-target} illustrates the process visually.
In this example, a large retailer has sales data coming from
stores in different geographical regions and across different
time-periods.   Some version of the desired pipeline
has been built previously -- the top row
of the figure shows a chunk of data
for ``\val{US-Store-1}'' and ``\val{2019-Dec}'', 
and for this chunk there may already be a legacy script/pipeline from IT 
that produces a database table or a dashboard.
However, as is often the case, new chunks of data 
for subsequent time-periods or new stores
need to be brought on-board, which however
may have different formats/schema 
(e.g., JSON vs. CSV, pivoted vs. relational, 
missing/extra columns, etc.),
because they are from different point-of-sales systems
or sales channels.   Building a new pipeline manually for
each such ``chunk'' (shown in the second/third row
in the figure) is laborious, and especially 
challenging for less-technical users who may not have the skills to
build such pipelines from scratch.  Today these less-technical users
often have to submit a ticket, and wait until IT has the
bandwidth to serve their needs.

\begin{figure}[t]
\vspace{-12mm}
\hspace{0mm}\subfigure{ \label{fig:table-target}
        \centering
        \includegraphics[width=0.48\columnwidth]{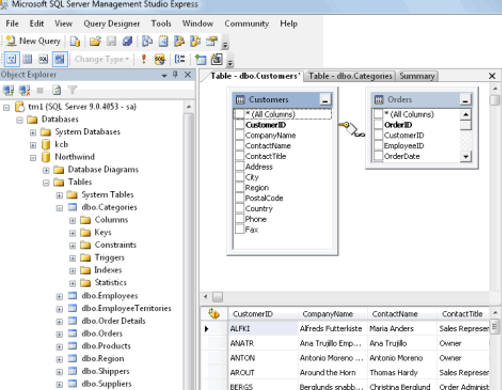}}	
\subfigure{ \label{fig:viz-target}
        \centering
       \includegraphics[width=0.48\columnwidth]{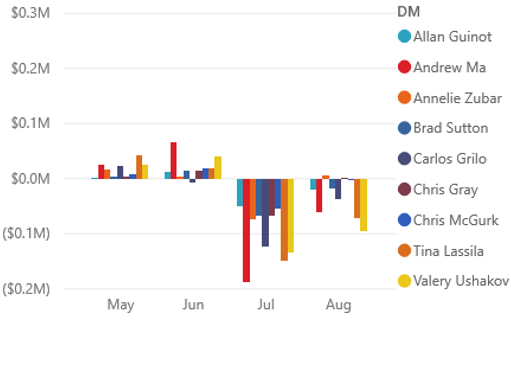}}	
       \vspace{-4mm}
       \caption{To trigger by-target synthesis, users only need to (Left): pick
an existing database table, right click and select
``\textsf{Append data to this table}'', or (Right): point to a
visualization, right click and select ``\textsf{Create a dashboard
like this}''.}
\label{fig:UI}
\vspace{-6mm}
\end{figure}

The aspirational question we ask, is whether pipelines can be 
synthesized automatically in such settings -- if 
users could point us to a ``target'' 
that schematically demonstrates
how the output should ``look like'',
as shown with green arrows in 
Figure~\ref{fig:by-target} that point to existing
database tables or visualizations. 
Concretely, ``targets'' can be specified like
shown in Figure~\ref{fig:UI}, where users could
right-click an existing database table and select the option to ``\val{append
data to the table}'', or right-click an existing visualization
and select ``\val{create a dashboard like this}'', to easily
trigger a pipeline synthesis process.

Unlike by-example synthesis, ``target'' used 
in this new paradigm  is only
as a fuzzy illustration of user intent. 
 Surprisingly, we show that this seemingly
imprecise specification is in fact often sufficient to 
uniquely determine the desired pipeline --
our insight is that
implicit constraints such as FDs and Keys discovered from
the target table are often sufficient to constrain the space 
of possible pipelines. This is a key
property overlooked thus-far by existing work, which we argue
can be the key to make pipeline-synthesis practical
(because fuzzy ``targets'' are a lot easier for users to provide).

\textbf{Search and RL-based Synthesis.}
The problem of synthesizing multi-step pipelines is clearly
challenging, as the number of candidate pipelines
grows exponentially in the number of steps, which is
prohibitively large very quickly (reaching $10^{20}$ within 5 steps
on typical tables having 10 columns).\iftoggle{fullversion}{\footnote{When given two 10-column tables, exhaustively
enumerating all possible join-paths  (assuming two
columns can join at most) would require $O(10^2)$ column-pairs
from one table, and $O(10^4)$ for both tables. (Similarly,
a 3-column group-by and 1-column aggregation would also generate
$O(10^4)$ options).  Without considering other operators, a 5-step pipeline
would already translate to $O(10^4)^5 = 10^{20}$ total option.}
}
{
}

In order to make synthesis tractable, we formalize
the end-to-end synthesis as an optimization problem, 
and develop a search-based algorithm \sjs{} that
considers a diverse array of factors
to best prioritize search over the most promising 
candidates.

We also design a deep 
reinforcement-learning (DRL) based synthesis algorithm 
\sjrl{}, which ``learns'' to synthesize pipelines 
using large collections of real
pipelines. Drawing inspiration from the success of
using ``self-play'' to
train game-playing agents like 
AlphaGo~\cite{silver2017mastering} and 
Atari~\cite{mnih2013playing},  we use ``self-synthesis'' 
to train an agent by asking it to try to synthesize
real pipelines, and rewarding
it when it succeeds.   It turns out that the RL-based
synthesis can learn to synthesize fairly quickly,
and slightly outperforms hand-crafted search using \sjs.

\iftoggle{fullversion}{
We summarize our contributions in the paper as follows:
\begin{itemize}[leftmargin=*]
\item We propose a new by-target synthesis paradigm that relieves users from needing to provide a precise output table, as is typically required in standard by-example synthesis.
\item We formulate the by-target synthesis as an optimization problem with inferred FD and Key constraints to guide the search.
\item We develop search and RL algorithms for by-target that can accurately synthesize real GitHub pipelines at interactive speeds.
\end{itemize}
}
{
}


\begin{figure*}[t]
\vspace{-12mm}
    \centering
    \includegraphics[width=1.6\columnwidth]{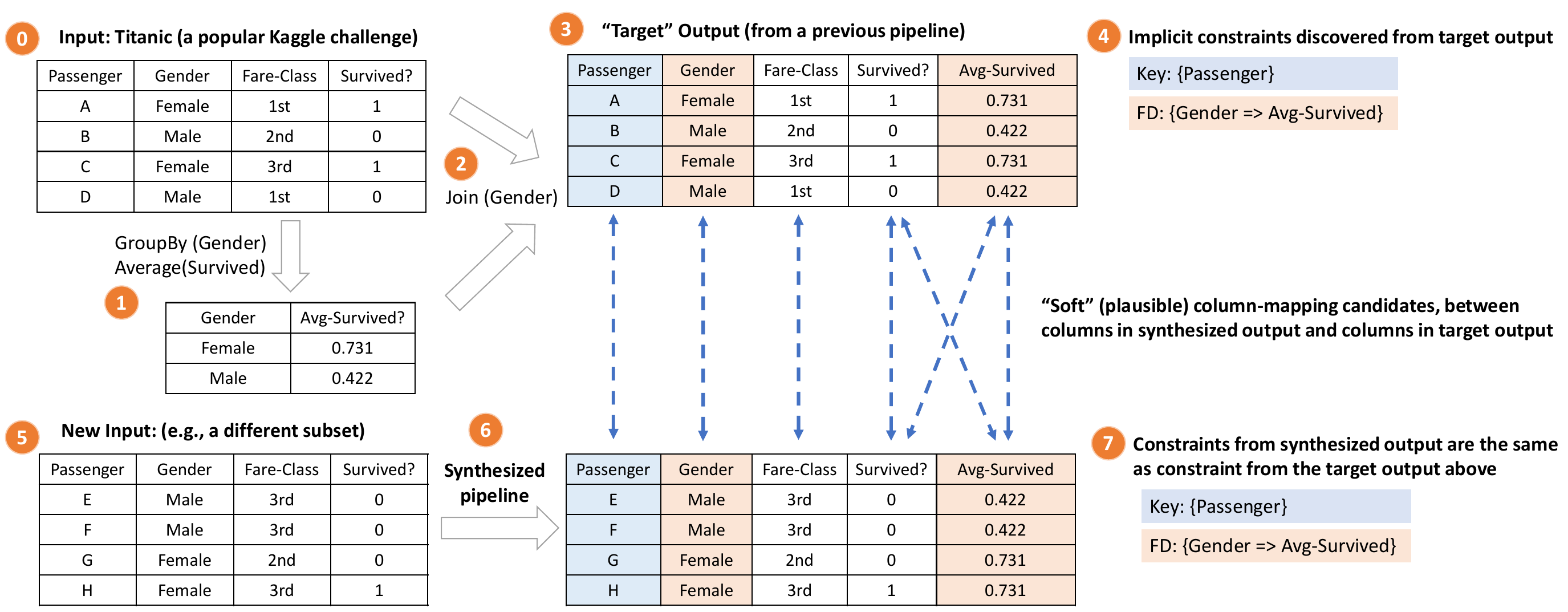}
   \vspace{-4mm}
    \caption{An example pipeline to show why ``by-target'' provides a
sufficient specification. Given (0) an input Titanic
table from Kaggle to predict
passenger survivals, a manually-authored pipeline performs
(1) a GroupBy on ``\val{Gender}'' to compute 
``\val{Avg-Survived}'' by ``\val{Gender}'',
and then (2) Joins it back on ``\val{Gender}''. 
Imagine that users give the output table (3) as the ``target'', 
we can discover constraints such as (4)
Key:\{``\val{Passenger}''\} and FD: \{``\val{Gender}''$\rightarrow$ ``\val{Avg-Survived}''\}.
(5) For a new input table (with a different set of passengers)
and given (3) as the ``target'', intuitively a correctly synthesized pipeline
in (6) should have the same FD/Key constraints that match
the ones from the target table (3), like shown in (7). }
    \label{fig:by-target-intuition}   
    \vspace{-4mm}
\end{figure*}

%% file: problem.tex
\section{Multi-Step By-Target Synthesis} 
\label{sec:problem}

We describe  the by-target synthesis problem in this section,
and we will start with preliminaries.

\vspace{-2mm}
\subsection{Preliminary: Pipelines and Operators}
\label{sec:preliminary}

\textbf{Data-pipelines.} 
Data pipelines are ubiquitous today, to transform raw data 
into suitable formats for downstream processing. 
Step (0)-(3) of Figure~\ref{fig:by-target-intuition} 
shows a conceptual pipeline using the \val{Titanic}
table as input, which is a popular Kaggle task to predict
which passengers survived~\cite{titanic-kaggle}.
The pipeline in this case performs
(1) a GroupBy on the \val{Gender} column to compute 
\val{Avg-Survived} by \val{Gender},
and then (2) a Join of the result with the input 
table on \val{Gender}, so that in (3) \val{Avg-Survived}
becomes a useful feature for predictions.

Today pipelines like this are built by both experts 
(e.g., developers and data-scientists)
and less-technical users (e.g., end-users in tools like Power Query
and Tableau Prep).

\begin{figure*}
\vspace{-12mm}
\hspace{0mm}\subfigure[A pipeline authored using Python Pandas]{ \label{fig:ex-pipeline-jupyter}
        \centering
        \includegraphics[width=0.8\columnwidth]{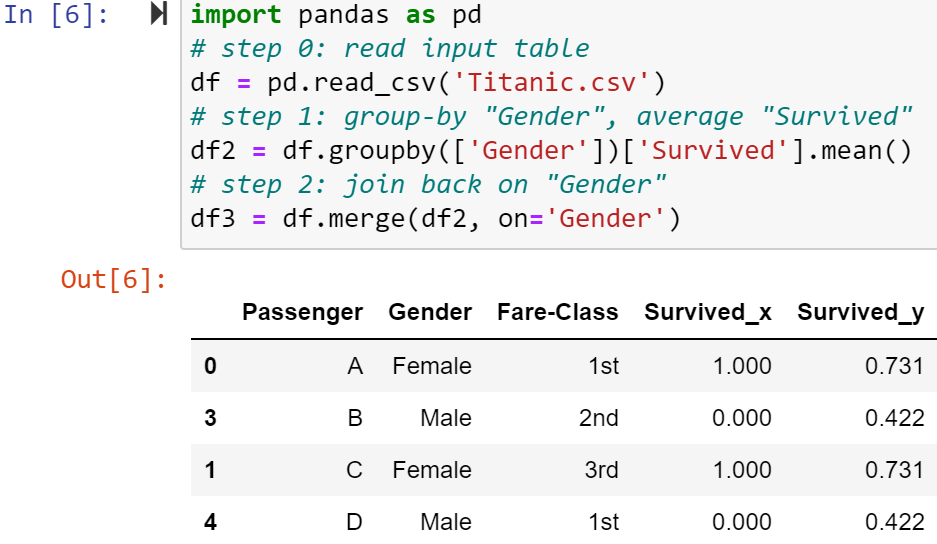}}	
\hspace{7mm}
\subfigure[A pipeline authored in visual drag-and-drop tool]{ \label{fig:ex-pipeline-tool}
        \centering
       \includegraphics[width=0.55\columnwidth]{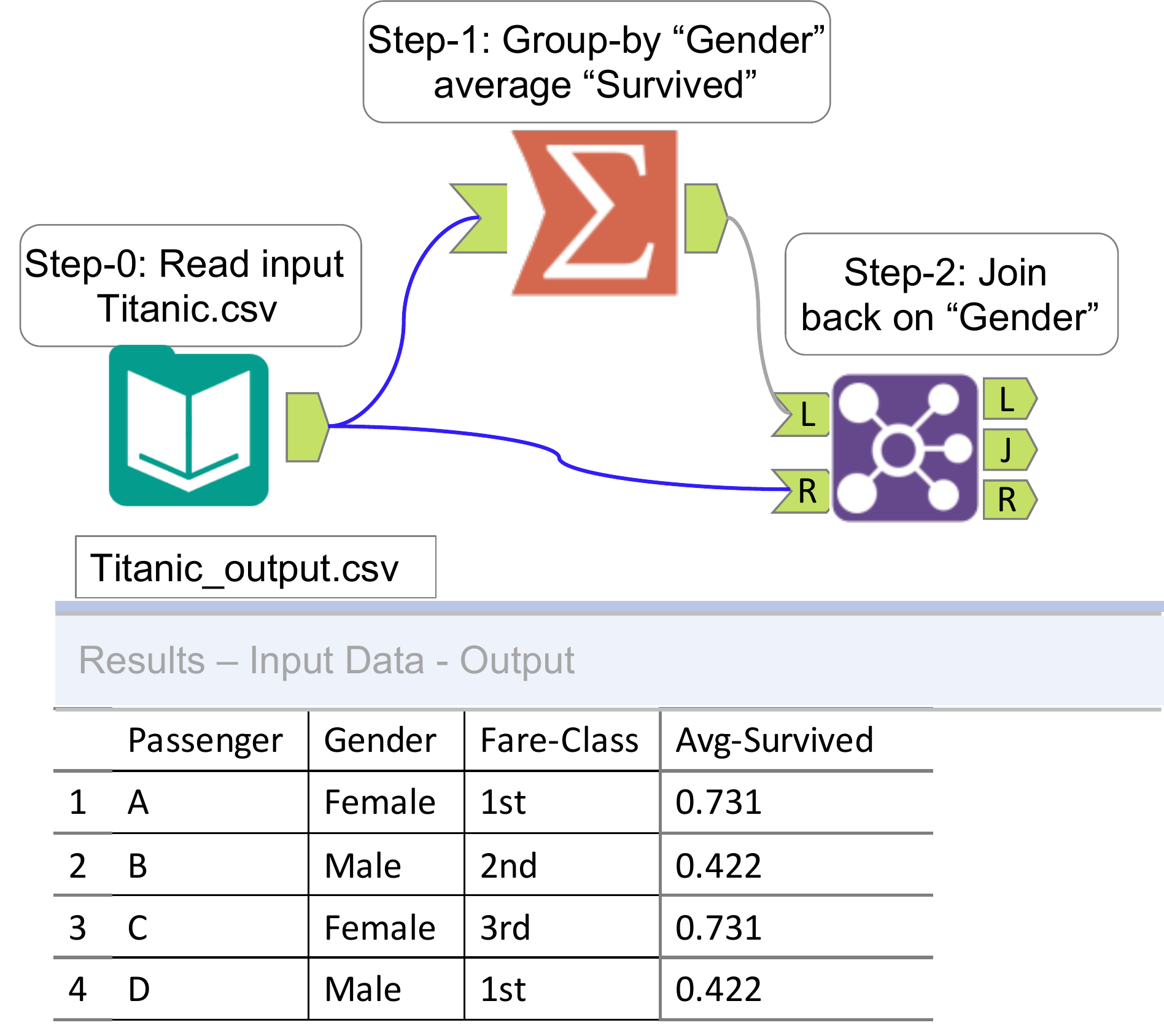}}
\vspace{-5mm}
\caption{Example pipelines corresponding to the same
steps in Figure~\ref{fig:by-target-intuition}. (a): A pipeline
built by data scientists using Python Pandas in a 
Jupyter Notebook. (b): The same pipeline built by 
less-technical users using visual drag-and-drop tools.}
\label{fig:ex-pipelines}
\end{figure*}

Expert users typically build pipelines using code/script,
with Pandas~\cite{pandas} in Python being particularly popular
for table manipulation. 
Figure~\ref{fig:ex-pipeline-jupyter}
shows an example pipeline written in Pandas
that corresponds to the same
steps of Figure~\ref{fig:by-target-intuition}. 
Today a lot of these pipelines
are written in Jupyter Notebooks~\cite{jupyter} and are publicly
available online. We crawled over 4M such notebooks on
 GitHub~\cite{auto-suggest}, from which
we can extract large quantities of real data pipelines.

Less-technical users also increasingly need 
to build pipelines themselves today, typically using
drag-and-drop tools 
(e.g., PowerQuery, Informatica, Azure Data Factory, etc.)
to manually specify pipelines step-by-step. 
Figure~\ref{fig:ex-pipeline-tool}
shows an example pipeline with the same
steps as Figure~\ref{fig:by-target-intuition}, but built
in a visual drag-and-drop tool, which are
more accessible to less-technical non-programmer users.

We note that the two pipelines
in Figure~\ref{fig:ex-pipelines} are equivalent, 
because they invoke the same sequence of \textit{operators} (a GroupBy
followed by Join). We introduce the notion of operators below.

\begin{figure}[t]
    \centering
    \includegraphics[width=1\columnwidth]{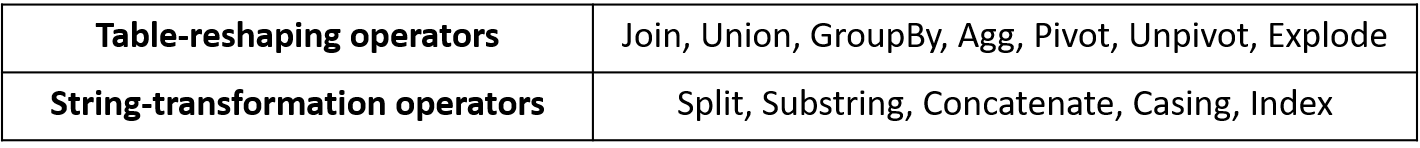}
    \vspace{-6mm}
    \caption{Operators considered in by-target synthesis.}
    \label{fig:DSL}
    \vspace{-5mm}
\end{figure}

\textbf{Operators.}
Conceptually, data-pipelines invoke
sequences of \textit{operators} that
broadly fall into two categories:

(1) \textit{Table-level operators}: e.g., Join, 
Union, GroupBy, Pivot, Unpivot, etc.
that manipulate tables.  A subset of these operators
are considered in SQL-by-example~\cite{qbo, sql-by-example}.

(2) \textit{String-level operators}: e.g., Split, 
Substring, Concatenate, etc., that perform 
string-to-string transformations. These operators 
are traditionally considered in 
transformation-by-example~\cite{HarrisG11, He2018TDE}.

In this work, we consider both classes of operators, since both are common
in pipelines. Figure~\ref{fig:DSL}
shows the operators we consider,
henceforth referred to as $\mathbf{O}$. 
\iftoggle{fullversion}
{
We give a description of these operators used in our DSL below.
\begin{itemize} 
\item Join: The Join operator is a standard 
operator in databases, which is also known as 
merge in Pandas~\cite{pandas-merge}. 
Automating the Join step, such as predicting likely parameters 
for Join (e.g., which columns to Join), is 
studied extensively in the literature 
(e.g.,~\cite{auto-suggest, rostin2009machine}).
\item GroupBy: The GroupBy operator is 
another a standard database operator, and
widely-used in table manipulation like Pandas~\cite{pandas-groupby}.
 Automating the GroupBy step by predicting
likely GroupBy columns is studied in~\cite{auto-suggest} (Section 4.2).
\item Aggregation: The Aggregation operator often 
follows GroupBy and is also widely used in libraries like Pandas~\cite{pandas-aggregate}. Predicting likely parameters for
Aggregation is also studied in~\cite{auto-suggest} (Section 4.2).
\item Pivot: The Pivot operator is a difficult 
table-level operator that restructures 
tables (turning values in a column into multiple columns)~\cite{pandas-pivot}. 
Predictive models for Pivot
have been built in~\cite{auto-suggest} (Section 4.3).
\item Unpivot: The Unpivot operator is the inverse of Pivot~\cite{pandas-unpivot}. 
Models for Unpivot have been built in~\cite{auto-suggest} (Section 4.4).
\item Union: The Union operator combines together two tables that have similar
schema, which is achieved using concat in Pandas~\cite{pandas-union}.
We implement a single-step predictive model 
for Union like~\cite{auto-suggest}. 
\item Split/Substring/Concat/Casing/Index: These 
string-level operators are quite standard and 
commonly used in many program languages
(e.g., in Python~\cite{python-string} and in C\#~\cite{csharp-string}).
We use these string operators in the same syntax as defined in prior
work~\cite{auto-join} (Section 3.2).
\end{itemize}
}
{
Because these
operators are fairly standard (e.g., the automation of 
individual operators are studied in depth in a 
prior work~\cite{auto-suggest}), we defer
descriptions of these operators to a full version of the paper~\cite{full}.
}

\textbf{Limitations.}
We note that expert users can write ad-hoc
\textit{user-defined functions} (e.g., any python code) in their
pipelines, which are unfortunately intractable for 
program-synthesis even in simple cases (e.g., PSPACE-hard for arithmetic functions)~\cite{das2010synthesizing, ellis2015unsupervised},
and are thus not considered in our work. Similarly, we do not consider
row-level filtering because it is also intractable in general~\cite{qbo}.

\vspace{-2mm}
\subsection{Problem: Multi-step By-target Synthesis}
\label{sec:target}
As illustrated in Figure~\ref{fig:by-target},
 in our pipeline synthesis problem, we are
given as ``target'' an existing table $T^{tgt}$ 
(e.g., a database table or a dashboard), 
generated from a pipeline $L$ on 
a previous batch of input 
tables $\mathbf{T}^{in} = \{T_1, T_2, \ldots \}$, written
as $T^{tgt} = L(\mathbf{T}^{in})$.

As is often the case, new data files, denoted by
$\widehat{\mathbf{T}}^{in} = \{\widehat{T}_1, \widehat{T}_2, \ldots \}$, 
have similar content but may have different schema and 
representations (e.g., because they come 
from a different store/region/time-period, etc.). Users would
want to bring 
$\widehat{\mathbf{T}}^{in}$ onboard,
but $L$ is no longer applicable, and often also
not accessible\footnote{\small End-users
wanting to build 
a ``similar'' pipeline  targeting an existing database-table/dashboard 
often do not have access to the original
legacy pipelines $L$ built by IT, due to 
discover-ability and permission issues. As such, to ensure generality, 
in this work we do not assume the original
$L$ to be available as reference to synthesize new 
pipelines (though we are clearly more likely to succeed
if the original $L$ is available).}.

In this work, we ask the aspirational question of whether
new pipelines can be automatically synthesized, if
users can point us to the new input files
$\widehat{\mathbf{T}}^{in}$ and the target $T^{tgt}$,
to schematically demonstrate what output from
a desired pipeline should ``look like''.
This by-target synthesis problem is defined as follows:
\begin{definition}
\label{def:problem}
In \textit{by-target pipeline-synthesis}, given 
input data $\widehat{\mathbf{T}}^{in}$, and 
a target table $T^{tgt}$ generated from related
input $\mathbf{T}^{in}$ that schematically demonstrates the
desired output, we need to synthesize a 
pipeline $\widehat{L}$ using a predefined set of operators $\mathbf{O}$, 
such that $\widehat{T}^{o} = \widehat{L}(\widehat{\mathbf{T}}^{in})$ 
produces the desired output.
\end{definition}

\textbf{Evaluate synthesized pipelines from by-target.}
Since one may worry that a target-table $T^{tgt}$
only provides a fuzzy specification of the synthesis problem,
we will start by discussing how a by-target synthesis system
can be systematically evaluated.

In traditional by-example synthesis
(e.g., SQL-by-example~\cite{qbo, sql-by-example}), a pair
of \textit{matching} input/output tables ($\widehat{\mathbf{T}}^{in}$, 
$\widehat{T}^{o}$) 
is provided as input to synthesis algorithms 
(even though in practice $\widehat{T}^{o}$ is hard to come by). 
In such a setting, evaluating 
a synthesized program $\widehat{L}$
 often reduces to a simple check of whether
the synthesized output $\widehat{L}(\widehat{\mathbf{T}}^{in})$ 
is the same as $\widehat{T}^{o}$.



\begin{figure}[t]
    \centering
    \includegraphics[width=1\columnwidth]{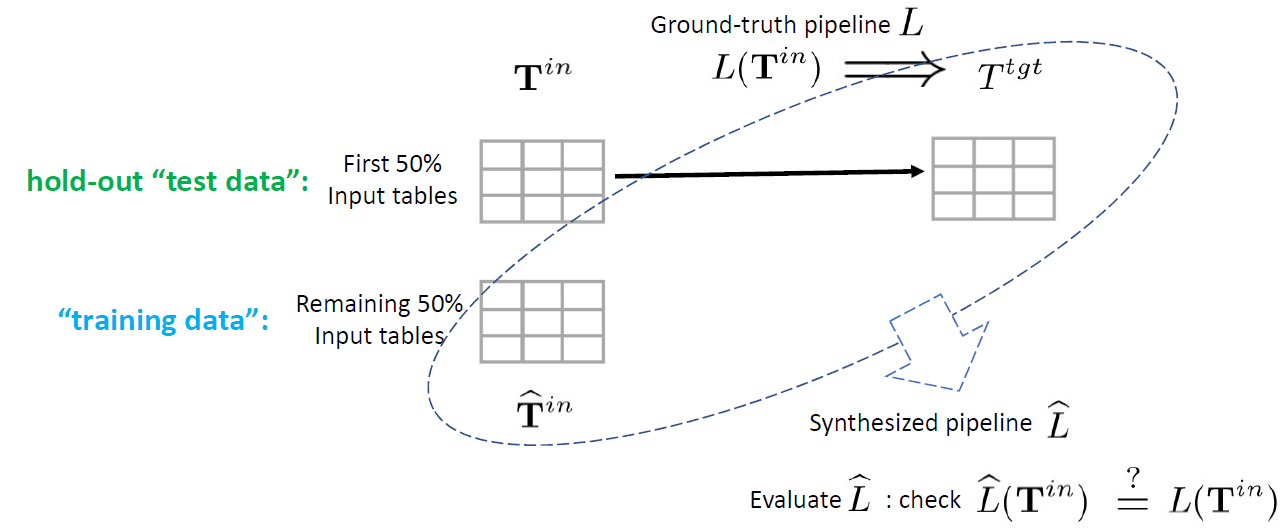}
    \vspace{-4mm}
    \caption{Evaluate by-target synthesis: Given a human-authored
pipeline $L$, we treat the first 50\% of input data for $L$ as
$\mathbf{T}^{in}$, to generate the target 
table $T^{tgt} = L(\mathbf{T}^{in})$. We then use the remaining 
50\% of input as $\widehat{\mathbf{T}}^{in}$, which
together with $T^{tgt}$, is fed into by-target synthesis to synthesize 
a new pipeline $\widehat{L}$. The correctness of $\widehat{L}$
can be verified on  $\mathbf{T}^{in}$ (held-out 
during synthesis), by checking whether 
$\widehat{L}(\mathbf{T}^{in}) \stackrel{?}{=} L(\mathbf{T}^{in})$.}
\vspace{-3mm}    
\label{fig:eval}
\end{figure}

In by-target synthesis, we are given as input 
a pair of \textit{non-matching} tables
($\widehat{\mathbf{T}}^{in}$, $T^{tgt}$),
for which the same evaluation does not apply. 
It turns out, however, that evaluation by-target
synthesis can be performed similarly, using what is
analogous to ``testing''/``training'' in 
Machine Learning. 

Specifically, as illustrated
in Figure~\ref{fig:eval}, for each 
real pipeline $L$ authored by humans, we
split the input tables used by $L$ 50\%/50\% into 
``testing'' and ``training''\footnote{\small When there are multiple
tables in a pipeline and Joins are required, we split the 
largest input table (which is fact-table-like) to ensure that
Joins do not produce empty results.}. We
treat the the first 50\% as if they are $\mathbf{T}^{in}$
in by-target synthesis, and
use the ground-truth pipeline $L$ to generate the target output
$T^{tgt} = L(\mathbf{T}^{in})$.  We then use
the remaining 50\% as if they are $\widehat{\mathbf{T}}^{in}$, 
and feed the non-matching pair
($\widehat{\mathbf{T}}^{in}$, $T^{tgt}$) as input
to by-target synthesis (circled in dash in Figure~\ref{fig:eval}),
so that a new pipeline $\widehat{L}$ can be synthesized. 
The correctness of the synthesized $\widehat{L}$
can be verified on the first 50\% data ($\mathbf{T}^{in}$),
which is held-out during synthesis, by checking whether
$\widehat{L}(\mathbf{T}^{in}) \stackrel{?}{=} L(\mathbf{T}^{in})$.
Note that because $\mathbf{T}^{in}$ is held-out
during synthesis (analogous to hold-out test-data in ML), 
and the original pipeline $L$ is also held-out,
the fact that we can ``reproduce'' a synthesized
$\widehat{L}$ that has the same effect as $L$ 
on the hold-out data $\mathbf{T}^{in}$ ensures that 
the synthesized $\widehat{L}$ from by-target
is indeed what users want.\footnote{\small Note that we do not 
require $\widehat{L}$ and $L$ to be identical at a syntactical-level, 
because there are often semantically equivalent ways to 
rewrite a pipeline (e.g., change of operator orders, or rewrite
using an equivalent sequence).}

\textbf{Is by-target a sufficient specification?}
Even though by-target synthesis can be systematically 
evaluated using a procedure analogous to train/test in ML,
one may still wonder whether 
a non-matching pair 
($\widehat{\mathbf{T}}^{in}$, $T^{tgt}$) in by-target synthesis
provides a sufficient specification for a desired pipeline 
to be synthesized. \iftoggle{fullversion}{(In any event, for cases where by-target
synthesis does not provide a sufficient specification 
such that desired pipelines cannot be found, all the failed
cases would still be reflected in the aforementioned evaluation).

}
{
}
We show that this seemingly
imprecise specification is in fact sufficient in most cases, 
by leveraging \textit{implicit constraints} that 
we can discover from $T^{tgt}$. We illustrate this 
using the following example. 

\begin{example}
Figure~\ref{fig:by-target-intuition} shows the conceptual
steps of a simple pipeline for the Titanic challenge~\cite{titanic}.
Like we discussed in Section~\ref{sec:preliminary},
this particular pipeline computes 
(1) a GroupBy on the \val{Gender} column to compute 
\val{Avg-Survived} by \val{Gender},
and then (2) a Join on \val{Gender} to bring \val{Avg-Survived} 
as an additional feature into the original input, like shown in (3).

In our setting of by-target synthesis, a different user
is now given a similar input table with
a different set of passengers like shown in (5).
Without having access to the original pipeline, 
she points to (3) as the target table to as a fuzzy demonstration
of her desired output, in order for by-target synthesis to produce 
the desired pipeline.  

Our key insight is that in such cases, the desired pipeline
can be uniquely determined, 
by leveraging \textit{implicit
constraints} discovered from the output table (3). 
Specifically, we can apply standard constraint-discovery
techniques (e.g.,~\cite{papenbrock2015functional}) 
to uncover two constraints shown in (4):
Key-column:\{``\val{Passenger}''\}, 
Functional-dependency (FD): \{``\val{Gender}'' 
$\rightarrow$ ``\val{Avg-Survived}''\}.

When table (5) is used as the new input and 
table (3) is used as the target, implicitly
we want a synthesized pipeline (6) to follow the same
set of transformations in the pipeline that produces (3), and 
as such the new
output using table (5) as input 
should naturally satisfy the same set of constraints.
Namely, if we perform a column-mapping between the table (3)
and table (6), we can see that the constraints discovered 
from these two tables, as shown in (4) and (7),
have direct one-to-one correspondence.
If we need to recreate these implicit constraints in table (3)
in a synthesized pipeline, 
it can be shown that the only pipeline
with the fewest steps to 
satisfy all these constraints is 
the aforementioned pipeline. (Others
would either miss one constraint, or
require more steps, which are less likely to be desired 
according to MDL and Occam's Razor~\cite{mdl}).
\footnote{We note that while the synthesized
pipeline in the example 
of Figure~\ref{fig:by-target-intuition} is the same as the original, there are many cases where synthesized pipelines are
different from the original, while still being
semantically equivalent. We defer this 
to Section~\ref{sec:comparison}. 
}
\end{example}

In summary, our key insight is that 
leveraging implicit constraints can sufficiently
constrain the synthesis problem. Our large-scale evaluation 
on real pipelines (Section~\ref{sec:experiments}) confirms that
most can indeed be successfully synthesized 
using the by-target paradigm.

\subsection{Synthesis Algorithm: Intuitive Sketch}
\label{sec:sketch}

\begin{figure}[t]
\vspace{-3mm}
    \centering
    \includegraphics[width=1\columnwidth]{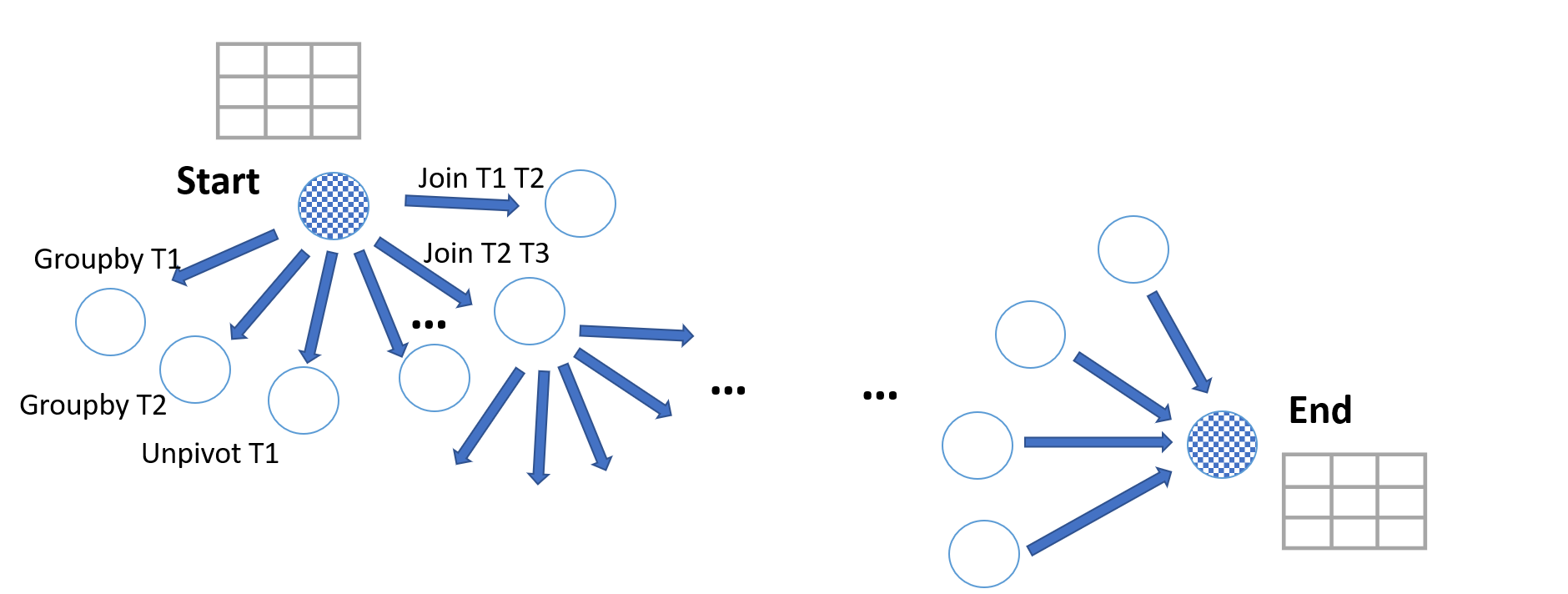}
    \vspace{-5mm}
    \caption{A search graph for synthesis: 
from the start-node (an empty pipeline) to the 
end-node (a synthesized pipeline),
each intermediate node represents a partial pipeline, and each edge 
represents the act of adding one operator, 
which leads to a new pipeline with one more operator.}
\vspace{-5mm}
    \label{fig:synthesis-sketch}
\end{figure}

We now give a sketch of how a synthesis
algorithm may look like before we formalize the problem.

Figure~\ref{fig:synthesis-sketch} gives
an intuitive illustration of the synthesis process. Each node 
here represents an intermediate state in the
synthesis process, which corresponds to a ``partial pipeline''.
The starting state (shown with a checkerboard pattern at the top-left)
corresponds to an empty pipeline $\widehat{L} = \{\}$, and 
the ending state (at bottom-right) 
corresponds to a final synthesized pipeline 
$\widehat{L} = \{O_1, O_2, \ldots  O_n\}$.

From each state representing a partial pipeline,
we can extend the partial pipeline by one additional
``step'' using some operator $O \in \mathbf{O}$ in
Figure~\ref{fig:DSL},
to move to a subsequent state.
For example, from the starting state $\widehat{L} = \{\}$, we can add
different instantiations of operators in $\mathbf{O}$
(e.g., different ways to apply GroupBy/Join/Pivot, etc.,
on given input tables),
which lead to different one-step pipelines (e.g., 
$\widehat{L} = \{\text{GroupBy(\val{table-1}, \val{column-1})}\}$).
This synthesis process can then be visualized as traversing the
search graph, until a satisfactory 
end-state is reached (e.g., satisfying 
all implicit constraints). 

It is clear from this intuitive sketch, however,
that the search space of possible pipelines is prohibitively large,
because (1) the number of possible pipelines grows exponentially
with the number of steps; and 
(2) even one individual step can be parameterized
in numerous ways -- e.g., 
a Join between two tables with $|C|$ columns each
can in theory use any of the $|C|^2$ column-pairs as the 
Join key (the
same is true for GroupBy/Pivot, etc.).

While we will defer a description of our solution to (1) above,
solving (2) is relatively straightforward 
because for each operator (e.g., Join), we
can leverage existing work (e.g.,~\cite{auto-suggest}) 
to accurately predict the most likely
way to parameterize the operator given
input tables (e.g., which columns to Join/GroupBy/Pivot, etc.).

\textbf{Predict Single-Operator Parameters.}
Conceptually, for each operator 
$O \in \mathbf{O}$, and given
input tables $T$, we need to predict the likelihood
of using parameter $p$ for $O$ in the context of $T$, 
written as $P_{T}(O(p))$.
For instance, for a Join between two given tables, we need
consider the characteristics of the tables 
to estimate which columns will likely join (which is a Join parameter);
similarly for Unpivot, we need to consider 
input tables and predict which subset of 
columns should Unpivot (also a parameter), etc.

For this reason, we build upon a prior technique called
Auto-Suggest~\cite{auto-suggest}, which learns from 
real data pipelines to
predict the likelihood of using parameters $p$ for each operator $O$ 
given input tables $T$, which is exactly $P_{T}(O(p))$.
In this work, we leverage~\cite{auto-suggest} and
treat these $P_{T}(O(p))$
as given, to better focus
on the end-to-end pipeline synthesis problem.
We refer readers to~\cite{auto-suggest}
for details of these single-operator predictions in the interest of space.

\textbf{Optimization-based formulation.}
Given the probabilistic estimates of operator parameters $P(O(p))$,
and the fact that we want to synthesize a pipeline that
can satisfy all implicit constraints (FD/Key),
we formulate the synthesis as an optimization problem.
Specifically, we want to find the ``most likely'' 
pipeline $\widehat{L}$ consisting
of a sequence of suitably parameterized operators
$\widehat{L} = \{O_1(p_1), O_2(p_2), \ldots \}$\footnote{\begin{small} While 
pipelines are in general directed acyclic graphs (DAGs), 
they can be serialized into sequences 
of invocations, thus the simplified notation. \end{small}},
by maximizing the joint probabilities of these operators $O_i(p_i)$, 
under the constraints that output from
$\widehat{L}$ should satisfy all implicit constraints. This problem,
henceforth referred to as PMPS (probability-maximizing
pipeline synthesis), can be written as follows:
\begin{small}
\begin{align}
\hspace{-1cm} \text{(PMPS)} \quad{} \argmax_{\widehat{L}}\quad{}\quad{} & \prod_{O_i(p_i) \in \widehat{L}}  { P(O_i(p_i)) }  \label{eqn:obj} \\ 
 \mbox{s.t.} ~~ & \text{FD}(\widehat{L}(\widehat{\mathbf{T}}^{in})) = \text{FD}(T^{tgt}) \label{eqn:fd} \\
 & \text{Key}(\widehat{L}(\widehat{\mathbf{T}}^{in})) = \text{Key}(T^{tgt}) \label{eqn:key} \\
 & \text{Col-Map}(\widehat{L}(\widehat{\mathbf{T}}^{in}), T^{tgt}) \label{eqn:map}
\end{align} 
\end{small}
The objective function 
in Equation~\eqref{eqn:obj}
states that we want to find the most likely pipeline $\widehat{L}$,
or the one whose joint probability of all single-step 
operator invocations is maximized.
Equation~\eqref{eqn:fd} and~\eqref{eqn:key} state
that when running the synthesized pipeline $\widehat{L}$
on the given input $\widehat{\mathbf{T}}^{in}$ to 
get $\widehat{L}(\widehat{\mathbf{T}}^{in})$, 
the FD/Key constraints discovered from $T^{tgt}$ should also be 
satisfied on $\widehat{L}(\widehat{\mathbf{T}}^{in})$. 
Finally Equation~\eqref{eqn:map} states that we should be able
to ``map'' columns from $\widehat{L}(\widehat{\mathbf{T}}^{in})$ to
$T^{tgt}$, with standard
schema-mapping~\cite{rahm2001survey}.

\begin{example}
We revisit Figure~\ref{fig:by-target-intuition}.
Using~\cite{auto-suggest},
we estimate the probabilities $P(O(p))$ of the two steps in the 
pipeline (GroupBy and Join) to be
0.4 and 0.8, respectively. Among all
other possible pipelines, this two-step pipeline maximizes
the joint probability (0.32) 
in Equation~\eqref{eqn:obj}, while satisfying all 
FD/Key/column-mapping constraints in 
Equation~\eqref{eqn:fd}-\eqref{eqn:map}, which is thus
the solution to PMPS.
\end{example}

%% file: algorithm-search.tex
\section{Search-based Auto-Pipeline}
\label{sec:algo}
This section describes our synthesis using
\sjs{}.

\subsection{A High-level Overview}
\label{sec:algo_overview}

As discussed in Section~\ref{sec:sketch}, 
at a high level the synthesis process can
be seen as traversing a large search graph
shown in Figure~\ref{fig:synthesis-sketch}.
Because each node corresponds to a partial-pipeline,
and each edge corresponds to the act of
adding one operator, each node that is $depth$-steps away from the
start-node would naturally correspond to a partial-pipeline 
with $depth$ number of operators/steps. 

Given the large search graph, it is natural
to explore only ``promising'' parts of the graph.
We first describe such a strategy in a meta-level synthesis algorithm
shown in Algorithm~\ref{algo:meta} below, which uses a form of 
beam search~\cite{ow1988filtered}.

\begin{algorithm}[t]
  \caption{\textsf{Synthesis}: A meta-level synthesis algorithm}
  \label{algo:meta}
\begin{small}
  \begin{algorithmic}[1]
 \Procedure{\textsf{Synthesis}}{$\widehat{\mathbf{T}}^{in}, T^{tgt}, \mathbf{O}$}
  \State $depth \gets 0, candidates \gets \emptyset$
  \State $S_{depth} \gets \{empty()\}$  \Comment{$\#$initialize an empty pipeline}  \label{line:meta_step_0}
  \While{$depth < maxDepth$} \label{line:meta_loop}
    \State{$depth \gets depth + 1 $}
    \ForEach{($L \in \mathcal S_{depth-1}$, $O \in \mathbf{O}$)}
      \State{$S_{depth} \gets S_{depth} \cup \textsf{AddOneStep}(L, O)$} \label{line:meta_add_one}
    \EndFor
    \State{$S_{depth} \gets \textsf{GetPromisingTopK}(S_{depth}, T^{tgt})$}  \label{line:meta_promising}
    \State{$candidates \gets candidates \cup \textsf{VerifyCands}(S_{depth}, T^{tgt})$}  \label{line:meta_verify}
  \EndWhile
  \State{return $\textsf{GetFinalTopK}(candidates)$} \label{line:meta_return}
\EndProcedure
  \end{algorithmic}
\end{small}
  \end{algorithm}

Algorithm~\ref{algo:meta} starts
by initializing $depth=0$ to indicate that we are
at the start-node in Figure~\ref{fig:synthesis-sketch}.
The variable $candidates$ stores all ``valid'' pipelines
satisfying the constraints in PMPS
(Equation~\eqref{eqn:fd}-\eqref{eqn:map}), and is
initialized as an empty set.
The variable $S_{depth}$ corresponds to all
pipelines with $depth$-steps that are
actively explored in one loop iteration, 
and at line~\ref{line:meta_step_0} we initialize it 
to a single placeholder empty-pipeline, because it is 
the only 0-step pipeline and we are still at the
start-node of the search graph.

From line~\ref{line:meta_loop}, 
we iteratively visit nodes that are $depth = \{1, 2, \ldots\}$
steps away from the start-node, which is equivalent
to exploring all pipelines with $\{1, 2, \ldots\}$ 
operators.   As we increment $depth$ in the loop,
we take all active pipelines from the previous iteration
with $(depth-1)$ steps, denoted by $S_{depth-1}$,
and ``extend'' each partial pipeline $L \in S_{depth-1}$
using one 
additional operator $O \in \mathbf{O}$, by
invoking \textsf{AddOneStep}($L, O$),
which is shown at line~\ref{line:meta_add_one}.
These resulting pipelines with $depth$-steps are
saved as $S_{depth}$.  
Because we cannot exhaustively explore all 
pipelines in $S_{depth}$, at line~\ref{line:meta_promising}, 
we select top-K most promising ones
from $S_{depth}$ by invoking \textsf{GetPromisingTopK}().
Among these top-K promising partial pipelines,
we check whether any of them already satisfy 
PMPS constraints using \textsf{VerifyCand}(),
and save the feasible solutions separately into $candidates$
(line~\ref{line:meta_verify}). This marks the end of one iteration.

We continue with the loop and 
go back to line~\ref{line:meta_loop}, where we increment
$depth$ by 1 and explore longer pipelines, until we find 
enough number of valid candidates, or we reach the maximum
depth, at which point we return the final 
top-K candidate pipelines
by invoking \textsf{GetFinalTopK}() (line~\ref{line:meta_return}). 

\textbf{Discussion.} While the key steps in our 
synthesis are sketched out in Algorithm~\ref{algo:meta},
we have yet to describe the sub-routines below:
\begin{itemize} 
\item \textsf{AddOneStep()} extends a partial pipeline 
$L$ using one additional operator $O \in \mathbf{O}$;
\item \textsf{VerifyCands()} checks whether pipelines
satisfy PMPS constraints, and if so marks them as final candidates;
\item \textsf{GetPromisingTopK()} selects the most promising
K pipelines from all explored pipelines with $depth$-steps;
\item \textsf{GetFinalTopK()} re-ranks and returns final
pipelines.
\end{itemize}
 
The first two sub-routines, \textsf{AddOneStep()} and 
\textsf{VerifyCands()}, are reasonably straightforward --
\textsf{AddOneStep()} adds one additional step
into partial pipelines by leveraging Auto-Suggest~\cite{auto-suggest} to find
most likely parameters for each operator,
while \textsf{VerifyCands()} checks for PMPS constraint 
using standard FD/key-discovery~\cite{buranosky2018fdtool, papenbrock2015functional} and
column-mapping~\cite{rahm2001survey}.
We will describe these two sub-routines in Section~\ref{sec:addonestep}
and~\ref{sec:verifycands}, respectively.

The last two sub-routines, \textsf{GetPromisingTopK()} and
\textsf{GetFinalTopK()}, are at the core of \sj{}, where a good design 
ensures that we can efficiently search promising
parts of the graph and synthesize successfully.
In Section~\ref{sec:diversity_search}, we 
will describe a search-based strategy to instantiate
these two sub-routines, and later in Section~\ref{sec:learning},
we will describe a learning-based alternative using RL.

\subsection{Extend pipelines by one step}
\label{sec:addonestep}
We describe the \textsf{AddOneStep()} subroutine in this section.\\
\textsf{AddOneStep($L, O$)} takes as input a $depth$-step partial
pipeline $L = \{O_1(p_1),$ $\ldots, O_{depth}(p_{depth}) \}$, and some operator $O$ (enumerated from
all possible  operators $\mathbf{O}$) that we want to add into $L$.
We leverages~\cite{auto-suggest},
which considers the characteristics of
intermediate tables in the partial pipeline $L$,
to predict the best parameter 
$p = \argmax_{p \in \mathbf{p}}{P(O(p)|L)}$ to use.
We use this predicted parameter $p$ to instantiate 
the new operator $O$, and use the resulting $O(p)$
to extend $L$ by one additional step, 
producing $L' = \{O_1(p_1),$ $\ldots O_{depth}(p_{depth}), O(p)\}$. 

Note that in general, for each operator $O$, there may be
more than one good way to parameterize $O$ 
(e.g., there may be more than one plausible GroupBy column, 
and more than one good Join column, etc.). So instead of
using only top-1 predicted parameter, for each $O$ 
we keep top-$M$ most likely parameters, 
which would produce $M$ possible pipelines
after invoking
\textsf{AddOneStep($L, O$)} for a given $L$ and $O$.

We use the following example to illustrate the process.
\begin{example}
\label{ex:extend}
We revisit the pipeline in Figure~\ref{fig:by-target-intuition}.
At step (0), we have one input table and an empty pipeline $L = \{\}$.
We enumerate all possible operators 
$O \in \mathbf{O}$ to extend $L$.

Suppose we first pick $O$ to be GroupBy. Intuitively we 
can see that \val{Gender} and \val{Fare-Class} columns are the
most likely for GroupBy (because among other things 
these two columns have 
categorical values with low cardinality). 
We leverage single-operator predictors
from~\cite{auto-suggest} -- in this case we use 
the GroupBy predictor (Section 4.2 of~\cite{auto-suggest}), 
which may predict that $P(GroupBy(\val{Fare-Class})|L) = 0.5$ and 
$P(GroupBy(\val{Gender})|L) = 0.4$ to be the most likely.
If we use $M=2$ or keep top-2 parameters for each operator,
this leads to two new 1-step pipelines
$L'_1 = \{GroupBy(\val{Fare-Class})\}$ and 
$L'_2 = \{GroupBy(\val{Gender})\}$.

The same process continues for other $O \in \mathbf{O}$. For
instance when we pick $O$ to be ``Pivot'',
we may predict that \val{Gender} and \val{Fare-Class} to be likely 
Pivot keys, so we 
get $L'_3 = \{Pivot(\val{Gender})\}$, $L'_4 = \{Pivot(\val{Fare-Class})\}$.

However, when we pick $O$ to be Join/Union, the
probabilities of all possible
parameters are 0, because no parameter is valid with 
only one input table in $L$. This changes when we have
more intermediate tables -- e.g., in a subsequent 
step marked as (1) in
Figure~\ref{fig:by-target-intuition}, a new intermediate
table is generated from GroupBy.
At that point, using~\cite{auto-suggest} we may predict a Join using  
\val{Gender} to be likely, while a Union is unlikely (because of the 
schema difference).
\end{example}

\subsection{Verify constraint satisfaction}
\label{sec:verifycands}
We now describe \textsf{VerifyCands()} in this section.  Recall that\\
\textsf{VerifyCands($S_{depth}, T^{tgt}$)} takes as input
a collection of pipelines $S_{depth}$ (the set of synthesized
pipelines with $depth$ steps), and check if
any $\widehat{L} \in S_{depth}$ satisfy all constraints 
 listed in Equation~\eqref{eqn:fd}-\eqref{eqn:map}
for Key/FD/column-mapping, in relation to the target table $T^{tgt}$.

\textbf{Column-mapping.} 
For column-mapping, we apply standard schema-mapping
techniques~\cite{rahm2001survey} to find possible column-mapping
between the target
table $T^{tgt}$, and the output table from a synthesized
pipeline $\widehat{L}(\widehat{\mathbf{T}}^{in})$, using a combination
of signals from column-names and column-values/patterns. 
\iftoggle{fullversion}
{

We will give an illustrative high-level example below, before describing
the details of the column-mapping algorithm we use (which is a variant
of standard schema-mapping~\cite{rahm2001survey}).
}
{
In the interest of space, we defer details of this
to a full version of the paper~\cite{full}, but we 
give an example below for illustration.
}

\begin{example}
\label{ex:col-mapping}
Consider a synthesized pipeline $\widehat{L}$ that
produces an output table shown at step (6) 
of Figure~\ref{fig:by-target-intuition}. Recall that our target
table $T^{tgt}$ is shown in step (3) -- for a synthesized $\widehat{L}$
to be successful, its 
output  $\widehat{L}(\widehat{\mathbf{T}}^{in})$ 
should ``cover'' all columns
in the target $T^{tgt}$, as required in 
Equation~\eqref{eqn:map}. As such, we need to establish
a column-mapping between the table in (3)
and the table in (6).

Using standard schema-mapping techniques, 
we find column-to-column mapping shown in 
Figure~\ref{fig:by-target-intuition}, using a combination
of signals from column-values, column-patterns, and 
column-names. 

It should be noted that our mapping
is \textit{not} required to be hard 1:1 mapping, but can be
\textit{``soft''} 1:N mapping. Like shown in  
Figure~\ref{fig:by-target-intuition}, 
the \val{Avg-Survived} column may
be mapped to both \val{Survived} and \val{Avg-Survived}
in the other table, since both share similar column-names
and values, and can be plausible mapping
candidates. In the end, so long as columns in $T^{tgt}$
can be ``covered'' by some plausible mapping candidates
in the synthesized result
$\widehat{L}(\widehat{\mathbf{T}}^{in})$,
this synthesized pipeline
$\widehat{L}$ is deemed to 
satisfy the column-mapping constraint in Equation~\eqref{eqn:map}.
\end{example}

\iftoggle{fullversion}
{
\textit{\underline{Algorithm details.}} We
now describe details of the column-mapping algorithm 
we use. Given two tables with $m$ and $n$ columns respectively,
we generate likely mappings between these columns as follows.
For all $m \times n$ column pairs, we first prune away column pairs of
different types (e.g., one is string while the other is number) from consideration.

For each of $(C_i, C_j)$ column pair that are string-valued,
we calculate the Jaccard similarity
 between the values of the two column, and consider the two
to have a likely mapping if Jaccard$(C_i, C_j)$ is over a threshold. 

For column pairs that do not meet the Jaccard threshold, 
we also calculate their average ``pattern distance'' to 
complement Jaccard that directly computes value overlap.
Specifically, we first ``normalize'' (denoted using function $\mathcal{N}$)
all [a-z] in a lowercase ``a'',
[A-Z] into an uppercase ``A'', and [0-9] into ``0'', 
so that we retain a ``patterns string'' for each value that 
abstracts away the specific values (e.g., $\mathcal{N}$(``John Doe'') = ``Aaaa Aaa'' and 
$\mathcal{N}$(``James Smith'') = ``Aaaaa Aaaaa'').

We then define relative pattern edit-distance (RPE) between two strings,
denoted by  $RPE(u, v)$, as the relative edit-distance 
between pattern-normalized $u$ and $v$,
or $RPE(u, v) = \frac{Edit(\mathcal{N}(u), \mathcal{N}(v))}{len(u) + len(v)}$.
For instance, the relative pattern edit-distance (RPE) between 
``John Doe'' and ``John Doe'' is $\frac{3}{19}$
(because the edit-distance between ``Aaaa Aaa''
and ``Aaaaa Aaaaa'' is 3, and the sum of their lengths is 19).

We calculate the minimum RPE (MRPE) between two columns 
$(C_i, C_j)$ as:
\[
MRPE(C_i, C_j) =\frac{1}{2}( \avg_{v_i \in C_i}{\min_{v_j \in C_j}{RPE(v_i, v_j)}}  +
 \avg_{v_j \in C_j}{\min_{v_i \in C_i}{RPE(v_i, v_j)}})
\].

For column pairs with little actual value overlap but high
pattern-overlap, e.g., $MRPE(C_i, C_j)$  is over a threshold, 
we consider them to be 
columns that likely map as well (e.g., so that two columns both
with people's names with similar value-patterns can be mapped, even though
they may not share same values between them).

Finally, for each of $(C_i, C_j)$ column pair that have numeric values,
we compute the range-overlap of $C_i$ and $C_j$,
defined as the intersection of the ranges of $C_i$ and $C_j$, 
over the union of the ranges.  We consider the two
as having a likely mapping if Range-overlap$(C_i, C_j)$ is over a threshold. 
}
{
}

\textbf{FD/Key constraints.} For FD/Key constraints, 
we again apply standard-constraint discovery
techniques~\cite{buranosky2018fdtool, papenbrock2015functional},
to discover FD/Key constraints from both the target
table $T^{tgt}$, and the output 
table $\widehat{L}(\widehat{\mathbf{T}}^{in})$ from a synthesized
pipeline $\widehat{L}$, in order to see if all FD/Key constraints 
from $T^{tgt}$ can be satisfied by $\widehat{L}$.
We use the example below to illustrate this.

\begin{example}
\label{ex:col-mapping}
Given a synthesized pipeline $\widehat{L}$ that
produces an output table shown at step (6) 
of Figure~\ref{fig:by-target-intuition}, and a target
table $T^{tgt}$ shown in step (3), 
we use constraint-discovery
to discover Key/FD constraints, which are shown in 
(7) and (4) for these two tables, respectively.
Given the soft column-mapping candidates 
shown in Figure~\ref{fig:by-target-intuition}, we can see that there exists
one column-mapping (with \val{Survived} $\leftrightarrow$
\val{Survived} and \val{Avg-Survived} $\leftrightarrow$ \val{Avg-Survived}),
under which all Key/FD constraints
from the target table (3) can be satisfied by the ones in table (6),
thus satisfying  Equation~\eqref{eqn:fd} and \eqref{eqn:key}.

Given that Key/FD/column-mapping 
have all been satisfied for the output in (6),
in \textsf{VerifyCands()} we can mark the corresponding
pipeline $\widehat{L}$ as a feasible solution to PMPS
(and saved in the variable $candidate$ in Algorithm~\ref{algo:meta}).
\end{example}

\subsection{A diversity-based search strategy}
\label{sec:diversity_search}
We now describe \textsf{GetPromisingTopK()} and
\textsf{GetFinalTopK()}, which are at the core of
 \sjs.

Recall that our goal is to solve the optimization
problem PMPS in Section~\ref{sec:sketch}, which
requires us to find a pipeline that can 
(1) maximize overall joint operator probabilities in the synthesized pipeline
(the objective function in Equation~\eqref{eqn:obj}), and 
(2) satisfy constraints in Equation~\eqref{eqn:fd}-\eqref{eqn:map}.

Because each candidate pipeline has already been checked for
constraint satisfaction (Equation~\eqref{eqn:fd}-\eqref{eqn:map}) 
in \textsf{VerifyCands()}, this makes
\textsf{GetFinalTopK()} easy as we only need to pick candidate
pipelines that maximize joint operator probabilities.
That is, for a synthesized pipeline
$\widehat{L} = \{O_1(p_1), O_2(p_2), \ldots \}$,
we can calculate its joint operator probabilities as 
$P(\widehat{L}) = \prod_{O_i(p_i) \in \widehat{L}}  { P(O_i(p_i)) }$,
where $P(O_i(p_i))$ are estimates from single-operator models
in~\cite{auto-suggest}. We can output a ranked list of top-K
pipelines, by simply ranking all candidate pipelines
using $P(\widehat{L})$.

On the other hand, the sub-routine \textsf{GetPromisingTopK()}
evaluates all $depth$-step pipelines currently explored,
where we need to find top-K promising candidates in order to prune
down the search space. We note that 
\textsf{GetPromisingTopK()} could not use the same strategy
as \textsf{GetFinalTopK()} by simply maximizing $P(\widehat{L})$,
because this may lead to pipelines that cannot satisfy
PMSP constraints (Equation~\ref{eqn:fd}-\eqref{eqn:map}),
resulting in infeasible solutions.

Because of this reason,  we design a diversity-based
strategy in \textsf{GetPromisingTopK()}, by not only 
picking partial pipelines that maximize 
the objective function in PMSP 
(Equation~\eqref{eqn:obj}), but also the ones
that satisfy the most number of FD/key/column-mapping constraints
in (Equation~\eqref{eqn:fd}-\eqref{eqn:map}). 
Specifically, given a budget of $K$ promising partial pipelines
that we can keep in $S_{depth}$, we consider a balanced set of
criteria by selecting $\frac{K}{3}$ pipelines from each of the three groups below:
\begin{itemize}[leftmargin=*]
\item[] (1) We select $\frac{K}{3}$ pipelines that have the highest 
overall probabilities $P(\widehat{L})$;
\item[] (2) We select $\frac{K}{3}$ 
pipelines whose output tables satisfy the most number of
FD/Key constraints in the target table;
\item[] (3) We select $\frac{K}{3}$ 
pipelines whose output tables can
``map'' the most number of columns in the target table.
\end{itemize}

We demonstrate this using an example below.
\begin{example}
\label{ex:diversity}
We continue with the Example~\ref{ex:extend} in 
Figure~\ref{fig:by-target-intuition}.
Suppose we have a budget of $K=3$ pipelines to keep.
Using the diversity-based search in \textsf{GetPromisingTopK()},
we can keep 1 pipeline each based on (1) probabilities, (2) key/FD constraints,
and (3) column-mapping, respectively.
For all 1-step pipelines considered in Example~\ref{ex:extend},
based on the criterion (1) we can see that the partial pipeline
$L'_1 = \{GroupBy(\val{Fare-Class})\}$ has the highest 
probability and will be selected, while based on the criterion (2) the pipeline
$L'_2 = \{GroupBy(\val{Gender})\}$ will
be selected as it satisfies an 
additional FD constraint found in the target table, etc.

Suppose that among all 1-step pipelines, we 
select the set $S_1 = \{L'_1, L'_2, L'_5\}$ as promising partial pipelines 
in \textsf{GetPromisingTopK()} given a $K=3$.
In the next iteration when we consider 2-step pipelines,
we will start from $S_1$ and consider different
ways to extend pipelines in $S_1$ using \textsf{AddOneStep()}.
We can see that extending $L'_2 \in S_1$ with a Join on \val{Gender}
yields a high probability pipeline satisfying all constraints,
which becomes a solution to PMPS. 

Note that in this example,
we prioritize our search on a promising set of 
$K=3$ pipelines at each depth-level,
without exploring all possible 1-step and 2-step pipelines.
\end{example}

\iftoggle{fullversion}{

\subsection{Additional Details.}
While Algorithm~\ref{algo:meta} outlines
key steps in our synthesis algorithm, there are a few
additional optimizations that we omitted due to space limit,
which we will describe here. The complete version of 
the pseudo-code for synthesis is shown in Algorithm~\ref{algo:full}.

\begin{algorithm}[t]
  \caption{\textsf{Synthesis}: a full version with optimizations}
  \label{algo:full}
\begin{small}
  \begin{algorithmic}[1]
 \Procedure{\textsf{Synthesis}}{$\widehat{\mathbf{T}}^{in}, T^{tgt}, \mathbf{O}$}
  \State $depth \gets 0, candidates \gets \emptyset$
  \State $S_{depth} \gets \{empty()\}$  \Comment{$\#$initialize an empty pipeline}  
  \State{\textsf{Relationalize}($T^{tgt}, \widehat{\mathbf{T}}^{in}$)}
  \While{$depth < maxDepth$} \label{line:full_relationalize}
    \State{$depth \gets depth + 1 $}
    \ForEach{($L \in \mathcal S_{depth-1}$, $O \in \mathbf{O}$)}
      \State{$S_{depth} \gets S_{depth} \cup \textsf{AddOneStep}(L, O)$}
    \EndFor
    \State{$S_{depth} \gets \textsf{GetPromisingTopK}(S_{depth}, T^{tgt})$} 
    \State{$candidates \gets candidates \cup \textsf{VerifyCands}(S_{depth}, T^{tgt})$}  
  \EndWhile
  \State{$\textsf{FineTune}(candidates)$} \label{line:full_finetune}
  \State{return $\textsf{GetFinalTopK}(candidates)$} 
\EndProcedure
  \end{algorithmic}
\end{small}
  \end{algorithm}

\textbf{Normalizing non-relational tables.}
Our algorithm operates under the assumption that 
tables are in standard relational forms, where
each column has consistent data values drawn from
the same underlying domain.  While this is almost always true
in traditional database settings, data tables
in the wild can have non-standard and non-relational
structures, such as Pivot tables~\cite{pivot-table}, which
can cause trouble in column-oriented reasoning (e.g.
column-mapping).

To optimize for scenarios where the target $T^{tgt}$
is non-relational (e.g., the result of a Pivot operation),
we perform the optimization 
(shown in line~\ref{line:full_relationalize} of
Algorithm~\ref{algo:full}) that 
normalizes $T^{tgt}$ and $\widehat{\mathbf{T}}^{in}$
into standard relational tables by invoking Unpivot, 
when it is predicted that an Unpivot is suitable (by again 
leveraging~\cite{auto-suggest}, in Section 4.4).

While this relationalization step can also be explored and searched
in the search graph in Figure~\ref{fig:synthesis-sketch} (e.g.
invoking Unpivot on $T^{tgt}$ and making Unpivot($T^{tgt}$)
the new target is equivalent to invoking
Pivot on the final table right before output),
we find this optimization to be worthwhile as it simplifies
the synthesis as it then reliably operate on tables that are
column-oriented.

\textbf{Fine-tuning of candidate pipelines.}
In line~\ref{line:full_finetune} of Algorithm~\ref{algo:full},
we perform a final-step fine-tuning of candidates, to ``salvage''
candidate pipelines that are really ``close'' to satisfying
all constraints on the target, but are still falling short a bit 
(e.g., missing an FD constraint or a Key constraint).

Specifically, we fine-tune final candidates with
a missing key column in relation to $T^g$, by adding a GroupBy
step on this missing key column. Similarly,
for candidate pipelines with missing FDs
we fine-tune by adding a GroupBy followed by a Join.

    \begin{figure}[t]
        \centering
        \subfigure[A case requiring substring.]{ 
            \includegraphics[width=\columnwidth]{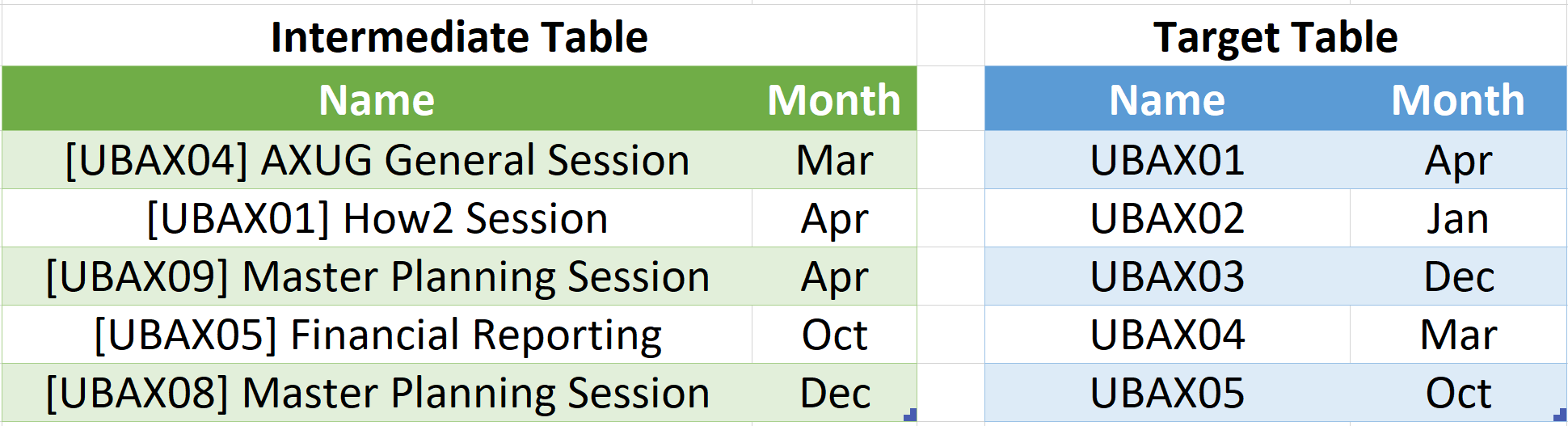}
            \label{fig:string-1}
            }\vspace{-3mm}
        \subfigure[A case requiring split and substring.]{ 
            \includegraphics[width=\columnwidth]{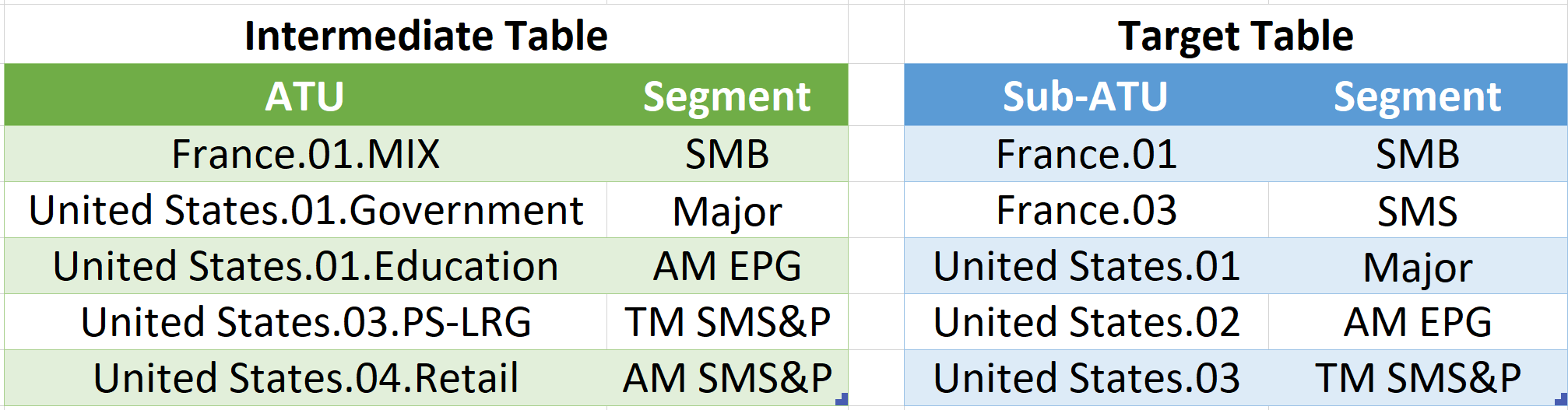}
            \label{fig:string-2}
            }
            \vspace{-3mm}
        \caption{Example cases requiring string-transformations.}
\vspace{-3mm}
        \label{fig:string}
    \end{figure}

In addition, we perform string-transformations in
the fine-tuning step, by comparing the intermediate 
result table from the partial pipeline, with the final target,
like shown in Figure~\ref{fig:string}.
In Figure~\ref{fig:string-1}, the left table is the
result table from a partial pipeline, and we see that a better
column-mapping can be created with the \val{name} column
in the target-table, if a Substring is invoked on the
\val{name} column in the intermediate table. 
Similarly, in Figure~\ref{fig:string-2}, a better
column-mapping can be created with the \val{Sub-ATU} column
in the target-table, if a Split followed by a Substring is invoked on the
\val{ATU} column in the intermediate table. 

We note that both of these cases can be readily addressed
using a prior work we developed for string-based 
program-synthesis~\cite{auto-join}. This module would take the left-table
and the right-table from Figure~\ref{fig:string} as input, and
produces a string program that can perform these desired transformations.
This string-transformation program can then be added as a final step
in the pipeline we synthesize. 

Because we directly leverage the code 
from~\cite{auto-join}, and the input/output
specification of this step is exactly the same,
we refer users to~\cite{auto-join} for details of 
how this step is implemented.
}
{
\textbf{Additional details.} 
While Algorithm~\ref{algo:meta} outlines
key steps in our synthesis algorithm, there are a few
additional optimizations, such as normalizing non-relational
input tables, and fine-tuning candidate
pipelines in final steps, etc. In the interest of
space we refer readers to
a full version of this paper~\cite{full} for additional details.
}

%% file: algorithm-learning.tex
\vspace{-2mm}
\section{Learning-based Auto-Pipeline}
\label{sec:learning}

In addition to the search-based synthesis, 
we also design a learning-based synthesis, 
which follows the exact same
steps in Algorithm~\ref{algo:meta}, except that
we replace the search-based heuristics in
\textsf{GetPromisingTopK} and \textsf{GetFinalTopK}
(in Section~\ref{sec:diversity_search}),
using deep reinforcement-learning (DRL) models.

\textbf{Learning-to-synthesize: key intuition.} 
At a high-level, our pipeline synthesis problem
bears strong resemblance to game-playing systems
like AlphaGo~\cite{silver2017mastering} and 
Atari~\cite{mnih2013playing}. 

Recall that in learning-to-play games like Go and Atari, agents
need to take into account game ``states'' they are in
(e.g., visual representations of game screens 
in Atari games, or board states in the Go game),
in order to produce suitable ``actions'' (e.g., pressing 
up/down/left/right/fire buttons in Atari, or placing a
stone on the board in Go) that are estimated to have the highest 
``value'' (producing the highest likelihood of winning).

In the case of pipeline synthesis, our problem
has a very similar structure. Specifically,
like illustrated in Figure~\ref{fig:synthesis-sketch},
at a given ``state'' in our search graph (representing a 
partial pipeline $L$), we need to decide suitable 
next ``actions'' to take -- i.e., among all possible ways
to extend $L$ using different operators/parameters,
which ones have the highest estimated
``value'' (giving us the best chance to synthesize
successfully).

Just like game-playing agents can be
trained via ``self-play''~\cite{mnih2013playing, silver2017mastering},
or by playing many episodes of games with win/loss outcomes 
to learn optimized ``policies''
 for games (what actions to take in which
states), we hypothesize that for pipeline-synthesis
an optimized synthesis ``policy'' may also be learned via ``self-synthesis'' --
namely, we could feed an RL agent with large numbers of real data pipelines,
asking the agent to synthesize pipelines by itself and assigning rewards
when it succeeds.

\textbf{Deep Q-Network (DQN).} 
Given this intuition, we set out to replace the 
search-based heuristics in
\textsf{GetPromisingTopK} and \textsf{GetFinalTopK},
using a particular form of reinforcement learning 
called Deep Q-Network (DQN)~\cite{mnih2013playing}, which uses a deep
neural network to directly estimate the ``value'' of a 
``state'', or intuitively how ``promising'' a partial pipeline is to
ultimately produce a successful synthesis.

More formally, like in Markov Decision Process (MDP),
we have a space of \textit{states} $\mathbf{S}$ where 
each state $s \in \mathbf{S}$ corresponds to a pipeline
$L(s) = \{O_1(p_1), O_2(p_2), \ldots, O_s(p_s) \}$, which
in turn corresponds to a node in our search graph in
Figure~\ref{fig:by-target-intuition}.

From each state $s \in \mathbf{S}$, we can take an 
\textit{action} $a \in \mathbf{A}$, which adds 
a parameterized operator to the pipeline $L(s)$
and leads to a new state $s'$ 
corresponding to the pipeline
$L(s') = \{O_1(p_1), O_2(p_2), \ldots,$ $ O_s(p_s), O_{a}(p_{a}) \}$.
Unlikely MDP, state transition in our problem
is deterministic, because adding $O_{a}(p_{a})$ to 
pipeline $L(s)$ uniquely determines a new pipeline. 

The challenge in our pipeline-synthesis problem, however, is
that the state/action space of one data-pipeline will be
different from another data-pipeline -- for example, 
the action of adding an operator ``Join(\val{Gender})''
in the pipeline of Figure~\ref{fig:by-target-intuition}
would not apply to other pipelines operating on different
input tables. This calls for a way to better
``represent'' the states and actions, so that learned synthesis 
policies can generalize across pipelines.

Because of this reason, we choose to use 
Deep Q-Network (DQN) to directly learn the 
\textit{value-function}~\cite{sutton2018reinforcement} of each state $s$,
denoted as $Q(s)$, which estimates the ``value'' of a state $s$, or
how ``promising'' $s$ is in terms of its chance of reaching
the desired target.

 \begin{figure*}[t]
 \vspace{-15mm}
     \centering
     \includegraphics[width=1.4\columnwidth]{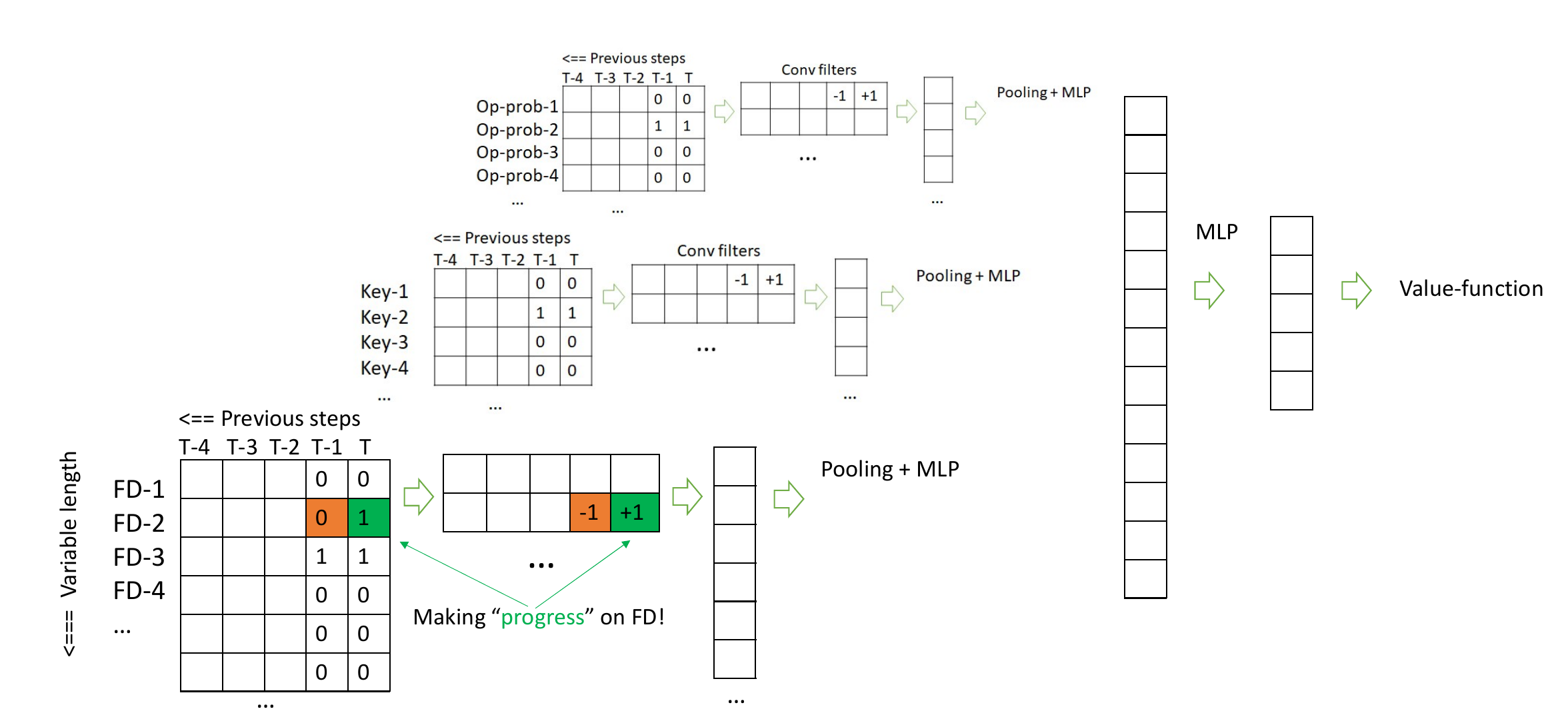}
     \vspace{-3mm}
     \caption{State representation for a partial pipeline at time-step $T$, using a convolution-based model.}
 \vspace{-3mm}
     \label{fig:model}
 \end{figure*}

\textbf{State representation and model architecture.} 
In order to represent states $\mathbf{S}$
of different pipelines in a manner that generalizes 
across different pipelines and tables, we need a 
representation that abstracts away the specifics of
each pipeline (e.g., which table column is used),
and instead encodes generic information important to by-target synthesis
that is applicable to all pipelines.

Recall that in our problem formulation PMPS, 
our end-goal is to synthesize a pipeline $\widehat{L}$ that
can produce all FD/Key constraints discovered
from the target $T^{tgt}$, while the operators
invoked in $\widehat{L}$ are plausible with
high estimated probabilities. As such,
in the representation we design,
we directly model these signals,
which are data/pipeline independent.

Given a pipeline $L_T$ 
with $T$ pipeline steps/operators, 
Figure~\ref{fig:model} shows the representation
we use to encode the state of $L_T$, including
FD/Key/operators/column-mapping, etc.
We will start with the representation for FD, which is
encoded using the matrix at lower-left corner
(other types of information are encoded similarly and will be described later). 

Recall that we discover FDs
from the target table $T^{tgt}$, and our goal is to synthesize
pipelines matching all these FDs. We arrange these
FDs in $T^{tgt}$ as rows in the matrix, and encode the FDs
satisfied by $L_T$ using the right-most columns
(marked with $T$), where a ``\val{0}'' entry indicates that
this corresponding FD has not been satisfied by $L_T$ yet,
while a ``\val{1}'' indicates that FD has been satisfied.
Columns to the left correspond to FDs of pipelines from previous 
time-steps, with $T-1$, $T-2$ steps/operators, etc., 
up to a fixed number of previous steps.

This representation, together with a convolutional architecture,
has two benefits. First, this explicitly models historical information
(i.e., what FDs are satisfied from previous pipelines), 
so that with convolution filters, we can
directly ``learn'' whether adding an operator at $T$-th step
makes ``progress'' in satisfying more FDs. 
As a concrete example, Figure~\ref{fig:model} shows 
a convolutional filter that may 
be learned, which have an orange ``-1''
and a green ``+1''. This filter  would allow
us to check whether an increased number of FDs are satisfied
from time $(T-1)$ to $T$. In this example, \val{FD-2} is not
satisfied at the $(T-1)$ step but is now satisfied at the
$T$ step, as marked by an orange ``\val{0}'' and green ``\val{1}''
in the matrix.
Applying a dot-product between the matrix and 
the example conv-filter, will yield
 (0*(-1) + 1*(+1)) = 1 on this portion of data,
 which can be intuitively seen as signaling ``positive-progress''
in covering more FDs from 
time $(T-1)$ to $T$. Observe that in comparison, 
from time $(T-1)$ to $T$, both \val{FD-1} (being 0/0 before/after) and \val{FD-3} (being 1/1 before/after) will get a 0 from
applying a dot-product with this filter, indicating 
no progress for these two FDs.

In typical computer-vision tasks where convolutional architectures
are applied, many conv-filters are stacked together
to learn visual features (e.g., circles vs. lines). 
We apply similar conv-filters in our synthesis
problem, which interestingly 
learn to observe local features like whether a pipeline-step
is making progress in FD/key/mapping, etc.

A benefit of using this representation with a convolutional architecture
is its ability to represent
pipelines with varying numbers of FDs/Keys, etc., because we can
set the number of rows in the matrix as 
the maximum number of FDs across all pipelines,
and conveniently ``pad'' rows not used for a specific pipeline as ``0''s
(which will always result in 0 irrespective of the conv-filters used).

In addition to FD, other types of information
(e.g., Key constraints, operator probabilities, column-mapping)
can be modeled similarly using the same matrix-representation
and conv-filter architecture, as shown in the top part
of Figure~\ref{fig:model}.
These representations are then fed into pooling and MLP
layers, before being concatenated and 
passed into additional layers to produce
a final function-approximation of $Q(s)$.

\textbf{Training via prioritized experience-replay.} 
We now describe our approach to train this model
to learn the value-function $Q(s)$,
using ``self-synthesis'' of real data pipelines harvested from GitHub. 

Similar to~\cite{auto-suggest},
we crawl large numbers of Jupyter notebooks on GitHub,
and ``replay'' them programmatically
to re-create real data pipelines, denoted by $\mathbf{L}$. 
We then train a reinforcement-learning agent to learn-to-synthesize
$\mathbf{L}$, by using Algorithm~\ref{algo:meta}
but replacing  \textsf{GetPromisingTopK} 
and \textsf{GetFinalTopK} with learned $Q(s)$
(i.e., picking top-K pipelines with the highest $Q(s)$).

We start with a $Q(s)$ model initialized using random weights. 
In each subsequent episode, we sample a real 
pipeline $L  = \{O_1(p_1),$ $O_2(p_2),$ $\ldots, O_n(p_n)\} 
\in \mathbf{L}$ and try to synthesize $L$ using the current $Q(s)$ and Algorithm~\ref{algo:meta}. If we successfully synthesize this $L$, 
we assign a reward of +1 for all previous states traversed by $L$
in the search graph -- that is, for all $i \in [n]$, 
we assign $Q(s_i) = +1$ where 
$s_i = \{O_1(p_1), \ldots, O_i(p_i)\}$\footnote{\small This corresponds to not
discounting rewards for previous steps, which is 
reasonable since we typically have only
around 10 steps.}. 
For all remaining states $s'$ traversed that do not lead to 
a successful synthesis, we assign $Q(s') = -1$.
By training the value-function $Q(s)$ using immediate
feedback, our hope is that
an optimized synthesis policy can be learned quickly that can take
into account diverse factors (operator probabilities  and 
various constraints).

We use \textit{prioritized experience 
replay}~\cite{schaul2015prioritized},
in which we record all $(s, Q(s))$ pairs in an internal memory M
and ``replay'' events sampled from M to update the model.
(This is shown to be advantageous because of its data efficiency, 
and the fact that events sampled over many episodes have
weak temporal correlations~\cite{lin1993reinforcement}). 
We train $Q(s)$ in an iterative manner
in experience replay. In each iteration, we use the $Q(s)$ from the
 previous iteration to play self-synthesis and collect a fixed $n$ number
of $(s, Q(s))$ events into the memory $M$. We use~\cite{schaul2015prioritized} to sample 
events in $M$ to update 
weights of $Q(s)$, and the new $Q'(s)$ will then be used 
to play the next round of self-synthesis.

In our experiments, we use $n=500$, and find the model to converge
quickly with 20 iterations. 
We also observe a clear benefit of using RL over standard 
supervised-learning (SL), because in RL we get to learn
from immediate positive/negative feedback
tailored to the current policy, which tends to be more informative
than SL data collected over fixed distributions.
\iftoggle{fullversion}
{

Additional details on the parameters used in our model: 
we use MSE error for learning the value-function $Q(s)$, with
adam optimizer, and mini-batch size 30. We track 5 previous
steps (so the matrix-width in Figure~\ref{fig:model} is 5). 
We use 3 types of 32 convolutional filters (with height being 3/4/5
and width being the same as the number of previous steps),
in each of the convolutional sub-models 
(for FD/Key/operators/column-mappings). 
In each sub-model, following the conv-filter layers,
we use 2D max-pooling and average-pooling, which are 
then concatenated together for the next layer.  Our two
final MLP layers have 32 and 8 units, respectively.
}
{
\footnote{\small Additional details
of our model can be found in a full paper~\cite{full}.} 
} 

%% file: experiment.tex
\vspace{-2mm}
\section{Experiments}
\label{sec:experiments}
We evaluate different pipeline synthesis algorithms
by both success rates and efficiency.
All experiments were performed on a Linux VM from a
commercial cloud, with 16 virtual CPU
and 64 GB memory. Variants of \sj{}
are implemented in Python 3.6.9.

\subsection{Evaluation Datasets}
\label{sec:dataset}
We created two benchmarks of data pipelines to evaluate the
task of pipeline synthesis, which have been made publicly
available\footnote{Our benchmark data is publicly available at: \url{https://gitlab.com/jwjwyoung/autopipeline-benchmarks}} to facilitate future research.

\textbf{The GitHub Benchmark.}
Our first benchmark, referred to as \textsf{GitHub},
consists of real data pipelines
authored by developers and data scientists, which we
harvested at scale from GitHub. Specifically, we crawled
Jupyter notebooks from GitHub, and
replayed them programmatically on corresponding data sets
(from GitHub, Kaggle, and other sources) to reconstruct the
pipelines authored by experts, in a manner similar
to~\cite{auto-suggest}.
We filter out pipelines
that are likely duplicates (e.g., copied/forked from other pipelines),
and ones that are trivially small (e.g., input tables have
less than 10 rows). These human-authored
pipelines become our ground-truth for by-target synthesis.

We group these pipelines based on \textit{pipeline-lengths},
defined as the number of steps in a pipeline. Longer pipelines are intuitively
more difficult to synthesize, because the space of possible
pipelines grow exponentially with the pipeline-length.
  For our synthesis benchmark, we randomly sample
100 pipelines of length
\{1, 2, 3, 4, 5, [6-8], 9+\}, for a total of 700 pipelines.



\textbf{The Commercial Benchmark.}
Since there are many commercial systems that also
help users build complex data pipelines
(e.g., vendors discussed
in Section~\ref{sec:intro}),
we create a second benchmark referred to as \textsf{Commercial},
using pipelines from commercial vendors.
 We sample 4 leading
vendors~\footnote{\small Alteryx~\cite{alteryx}, SQL Server Integration Services~\cite{ssis},
Microsoft Power Query~\cite{powerbi},
Microsoft Azure Data Factory~\cite{adf}}, and manually
collect 16 demo pipelines from official tutorials of
these vendors, as ground-truth pipelines for synthesis.

Recall that these commercial tools help users build pipelines
step-by-step (via drag-and-drop) -- with this benchmark
we aim to understand what fraction of pipelines from standard
commercial use cases (ETL and data-prep)
can be automated using \sj.

Note that for learning-based
\sj, we use models trained on the \textsf{GitHub} pipelines
to synthesize pipelines from the  \textsf{Commercial} benchmark,
which tests its generalizability.

\begin{table}
\centering
    \caption{Characteristics of pipeline synthesis benchmarks.}
    \vspace{-3mm}
    \label{tab:benchmarks}
\begin{small}
\begin{tabular}{lrrrr}
    \toprule
       Benchmark & \# of pipelines & \begin{tabular}{@{}c@{}}avg. \# of \\ input files \end{tabular}    & \begin{tabular}{@{}c@{}}avg. \# of \\ input cols \end{tabular} & \begin{tabular}{@{}c@{}}avg. \# of \\ input rows \end{tabular}  \\
        \midrule
        \textsf{GitHub}  & 700     & 6.6 & 9.1 &   4274     \\
        \midrule
        \textsf{Commercial}   & 16     & 3.75 & 8.7 & 988     \\
    \bottomrule
    \end{tabular}
\end{small}
\end{table}

\iftoggle{fullversion}{
\begin{figure}[h]
  \includegraphics[width=\columnwidth]{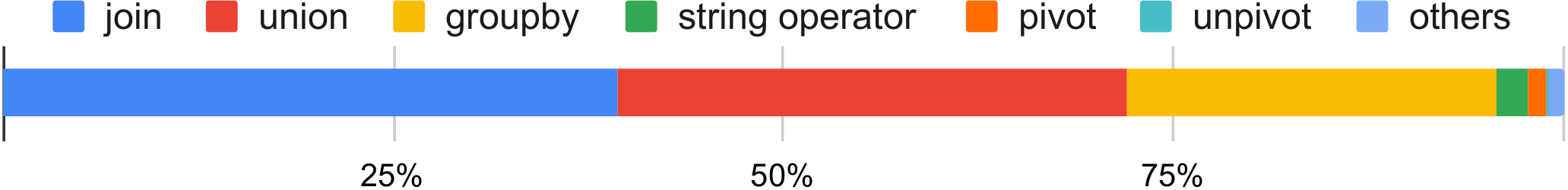}
  \vspace{-4mm}
  \caption{Operator distribution of pipeline benchmarks.}
  \label{fig:distributions}
  \vspace{-4mm}
\end{figure}
}
{
}

\subsection{Methods Compared}
Because ``by-target'' is a new paradigm
not studied in the literature before,
we compare \sj{} with
methods mostly from the ``by-example'' literature.

$\bullet$ \textbf{SQL-by-example}~\cite{sql-by-example}. This
recent ``by-example'' approach synthesizes SQL queries
by input/output tables.  Like other
``by-example'' approaches, \textsf{SQL-by-example}
requires users to provide an \textit{exact} output-table matching
the given input-tables. In order to
make it work, we provide the exact output of
the ground-truth pipelines to \textsf{SQL-by-example}.
We use the author's
implementation~\cite{sql-by-example-github}, and set a timeout
of 3600 seconds per pipeline. For cases where this method fails
due to timeout we give it another try using small input tables with
5 sampled rows only.

$\bullet$ \textbf{SQL-by-example-UB}~\cite{sql-by-example}.
Because \textsf{SQL-by-example} frequently
times-out on large input tables, we analyze its
theoretical upper-bound of ``coverage'', based on the operators
it supports in its DSL (Join, Aggregation, Union, etc.).
If all operators used in a benchmark pipeline are included
in its DSL, we mark the pipeline as ``covered'' in this
theoretical upper-bound analysis.

$\bullet$ \textbf{Query-by-output-UB (QBO-UB)}~\cite{qbo}.
\textsf{Query-by-output} is another influential ``by-example'' approach that
synthesizes SQL by input/output tables.
Since its code is not available, we also evaluate its
theoretical upper-bound coverage based on its operators.

$\bullet$ \textbf{Auto-Pandas}~\cite{bavishi2019autopandas}.
\textsf{Auto-Pandas} is another by-example approach that
synthesizes Pandas programs instead of SQL.
We use the authors' implementation~\cite{Auto-Pandas-code},
and like in \textsf{SQL-by-Example} we
feed it with ground-truth output-tables
matching the given input tables so that it can function properly.

$\bullet$ \textbf{Data-Context-UB}~\cite{koehler2019incorporating}.
This recent work proposes to leverage
data context (including data values and schema)
to automate mapping. Like Query-by-output-UB we evaluate its  theoretical upper-bound coverage based on the operators it handles.

$\bullet$ \textbf{Auto-Suggest}~\cite{auto-suggest}.
Auto-Suggest is a recent approach that
automates \textit{single} table-manipulation steps (e.g., Join,
Pivot, GroupBy) by learning from Jupyter Notebooks
on GitHub. We use \textsf{Auto-Suggest} to synthesize
multi-step pipelines, by greedily finding top-K most
likely operators at each step.

$\bullet$ \textbf{\sj}. This is our proposed method.
We report results from three variants, namely the search-based
\sjs, the supervised-learning-based \sjsl,
and the reinforcement-learning-based \sjrl.

For learning-based methods,
we randomly sample 1000 pipelines with at least 2 steps
as training data that are completely disjoint
with the test pipelines -- specifically, we
not only make sure that the train/test pipelines have no overlap,
but also the data files used by the train/test pipelines have
no overlap (e.g., if a pipeline
using ``titanic.csv'' as input is selected in the test set, no pipelines
using an input-file with the same schema would be selected into training).
This ensures that test pipelines are completely unseen to learning-based
synthesizers, which can better test synthesis on new pipelines.
Because learning-based variants uses stochastic gradient descent
during training, we report numbers
averaged over 5 offline training runs with different
seeds.\footnote{It should be
noted that once a model is trained, its predictions are
deterministic.}

\subsection{Evaluation Method and Metric}
\textbf{Accuracy.} Given a benchmark of $P$ pipelines,
accuracy measures the fraction of pipelines that can
be successfully synthesized, or $\frac{\text{num-succ-synthesized}}{P}$.

\textbf{Mean Reciprocal Rank (MRR).}
MRR is a standard metric that measures the quality
of ranking~\cite{ir}.  In our setting,
a synthesis algorithm returns a ranked list of $K$ candidate pipelines
for each test case,
ideally with the correct pipeline ranked high (at top-1).
The \textit{reciprocal-rank}~\cite{ir} in this case is defined as
$ \frac{1}{rank}$, where $rank$ is the rank-position of the first
correct pipeline in the candidates
(if no correct pipeline is found then the
reciprocal-rank is 0). For a benchmark with $P$ test pipelines,
the \textit{Mean Reciprocal Rank} is the mean
reciprocal rank over all pipelines, defined as:
\begin{center}
    $\mathrm{MRR} = \frac 1 P \sum_{i=1}^{P} \frac 1 {rank_i}$
\end{center}
We note that MRR is in the range of [0, 1], with 1 being perfect
(all desired pipelines ranked at top-1).


\subsection{Comparison of Synthesis}
\label{sec:comparison}

\textbf{Overall comparisons.} Table~\ref{tab:overall-github}
and Table~\ref{tab:overall-commercial} show an overall comparison
on the \textsf{GitHub} and
\textsf{Commercial} benchmark, respectively, measured
using accuracy, MRR, and latency. We report average
latency on successfully-synthesized
cases only, because some baselines would fail to synthesize
after hours of search.

As can be seen from
the tables, \sj{} based methods
can consistently synthesize 60-70\% of pipelines
within 10-20 seconds across the two benchmarks,
which is substantially more efficient and effective than
other methods.

While our search-based
\sjs{} is already effective,
\sjrl{} is slightly better in terms of accuracy.
The advantage of \sjrl{} over \sjs{} is more
pronounced in terms of MRR, which is expected
as learning-based methods are better at
understanding the nuance in fine-grained ranking decisions
than a coarse-grained optimization objective
in our search-based variant (Equation~\eqref{eqn:obj}).
\iftoggle{fullversion}{
\footnote{On the GitHub benchmark, the advantage
of \sjrl{} over \sjs{} is statistically significant for MRR with p-value at
0.03, but not significant for Accuracy. On the
Commercial benchmark, the observed
advantage is not statistically significant
because the benchmark has only 16 pipelines (it was collected
manually).}
}
{
}

We note that because our input/output tables are from
real pipeline and are typically large
(as shown in Table~\ref{tab:benchmarks}),
existing by-example synthesis methods like
\textsf{SQL-by-Example} frequently timeout
after hours of search, because their search methods are
exhaustive in nature. We should also note that
even the theoretical upper-bound coverage of existing by-example methods
(based on their DSL) are substantially smaller than
\sj, showing the richness of the operators supported in \sj.

\begin{table*}
\vspace{-15mm}
\parbox{.45\linewidth}{
\centering
    \caption{Results on the \textsf{GitHub} benchmark}
    \vspace{-3mm}
    \label{tab:overall-github}
\begin{small}
\begin{tabular}{lrrr}
    \toprule
        & Accuracy & MRR  & Latency (seconds) \\
        \midrule
        \sjs   & 76.6\%     & 0.724 & 18       \\
        \midrule
        \sjsl  & 73.7\%     & 0.583 & 20     \\
        \midrule
        \sjrl  & \textbf{76.9\%}     & \textbf{0.738} & 21       \\ \hline
        \midrule
        SQL-by-Example & 14.7\%     & 0.147 & 49 \\
        \midrule
        SQL-by-Example-UB & 56\%    & 0.56 & - \\
        \midrule
        Query-by-Output-UB & 15.7\%    & 0.157 & - \\
        \midrule
        Auto-Suggest & 29.7\%   & 0.297 & 11 \\
        \midrule
        Data-Context-UB & 43\% & 0.43 & -\\
        \midrule
        AutoPandas & 9 \% & 0.09 & 600\\
    \bottomrule
    \end{tabular}
\end{small}
}
\hfill
\parbox{.45\linewidth}{
\centering
    \caption{Results on the \textsf{Commercial} benchmark}
    \vspace{-3mm}
    \label{tab:overall-commercial}
\begin{small}
 \begin{tabular}{lrrr}
    \toprule
        & Accuracy & MRR  & Latency (seconds) \\
        \midrule
        \sjs   & 62.5\%     & 0.593 & 13       \\
        \midrule
        \sjsl   & \textbf{68.8\%}     & 0.583 & 14     \\
        \midrule
        \sjrl  & \textbf{68.8\%}     & \textbf{0.645} & 14       \\ \hline
        \midrule
        SQL-by-Example & 19\%     & 0.15 & 64 \\
        \midrule
        SQL-by-Example-UB & 37.5\%    & 0.375 & - \\
        \midrule
        Query-by-Output-UB & 18.8\%    & 0.188 & - \\
        \midrule
        Auto-Suggest & 25\%   & 0.25 & 13 \\
        \midrule
        Data-Context-UB & 25\% & 0.25 & - \\
        \midrule
        AutoPandas & 25\% & 0.25 & 34.5\\
    \bottomrule
    \end{tabular}
\end{small}
}
\end{table*}

\begin{figure*}
\minipage{0.3\textwidth}
  \includegraphics[width=\linewidth]{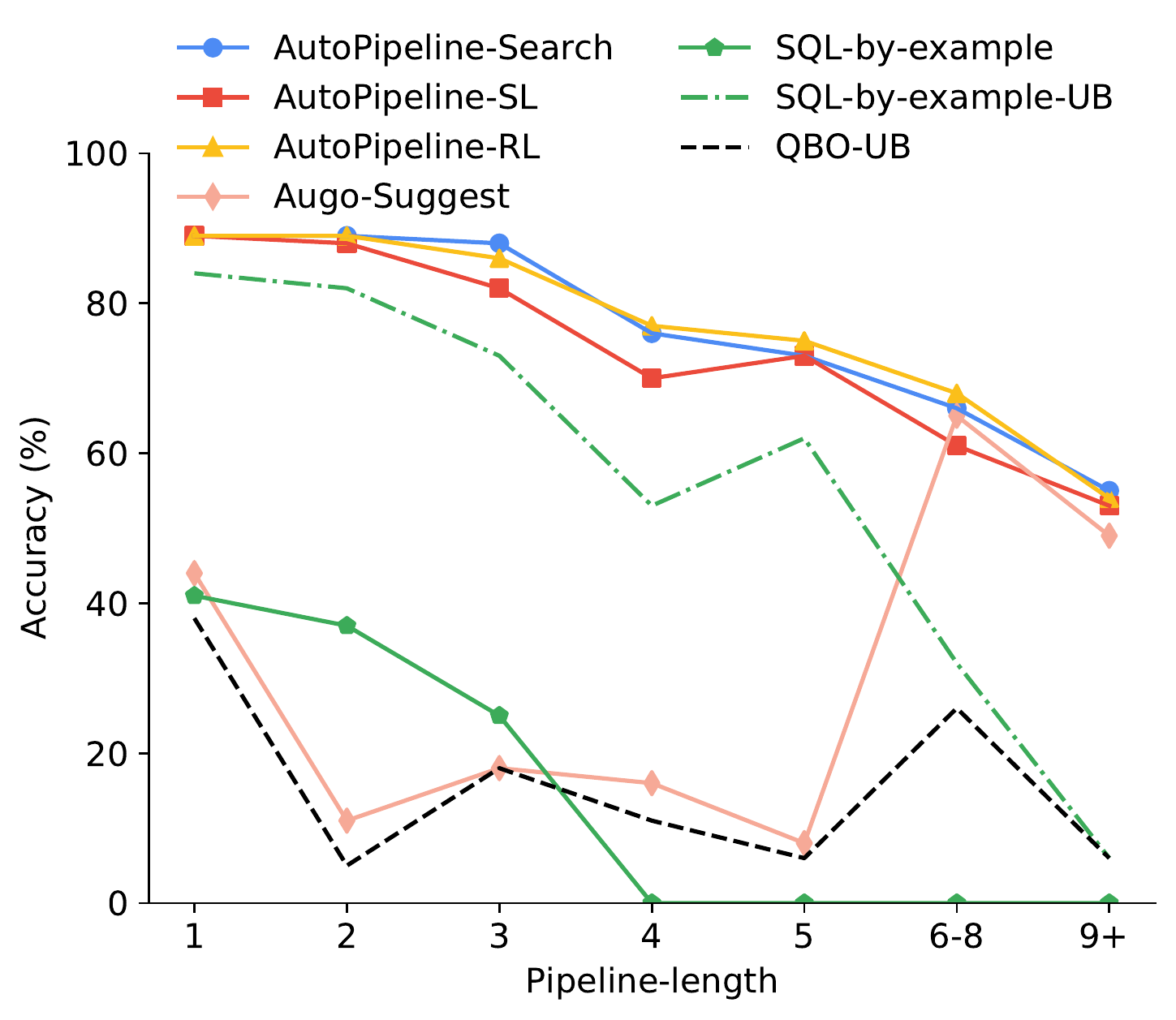}
  \vspace{-3mm}
  \caption{Accuracy results by pipeline-lengths on the \textsf{GitHub} benchmark.}
\vspace{-2mm}
    \label{fig:acc-comp}
\endminipage\hfill
\minipage{0.3\textwidth}
  \includegraphics[width=\linewidth]{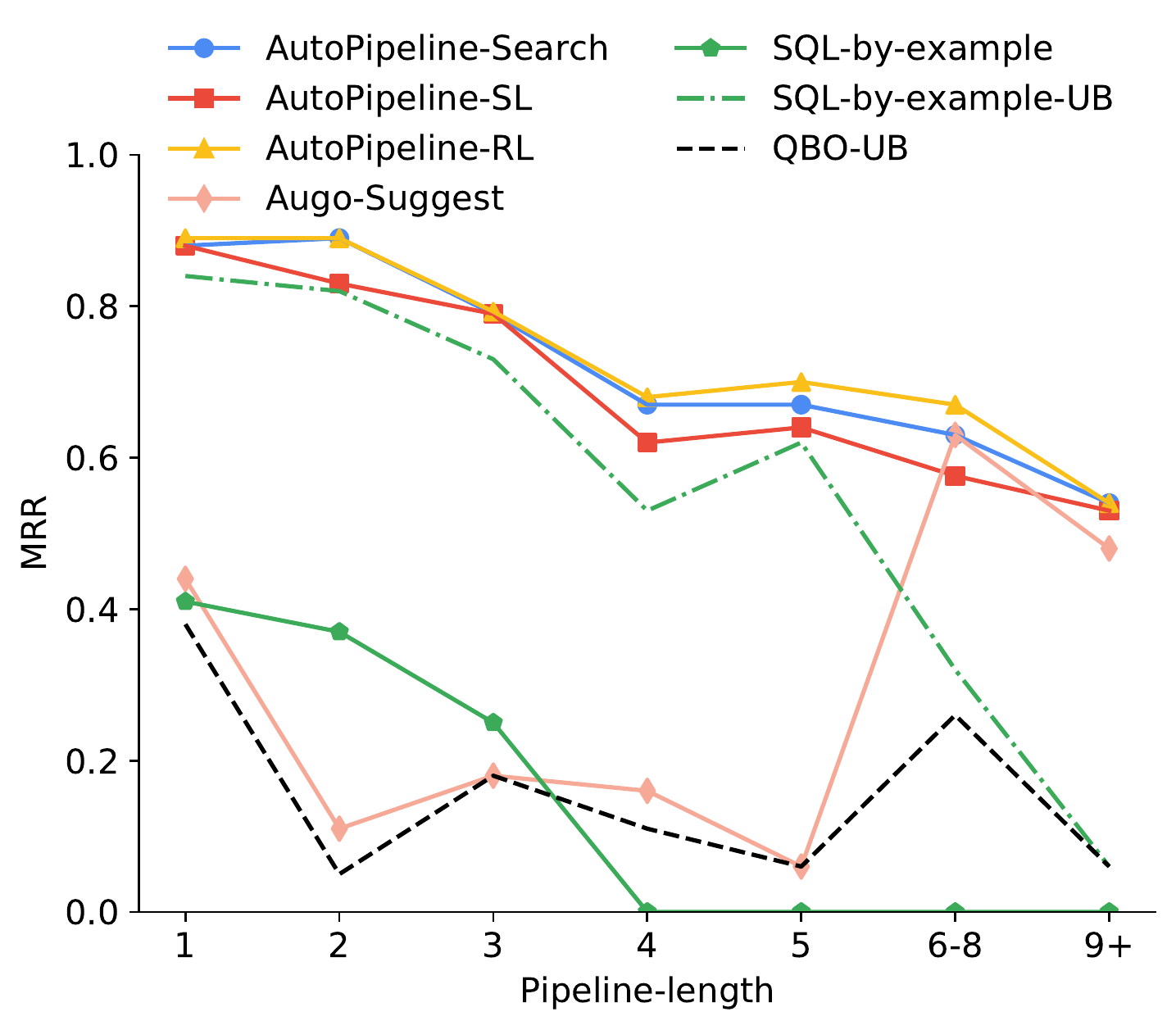}
  \vspace{-3mm}
  \caption{MRR results by pipeline-lengths on the \textsf{GitHub} benchmark.}
    \label{fig:mrr-comp}
\endminipage\hfill
\minipage{0.3\textwidth}%
  \includegraphics[width=\linewidth]{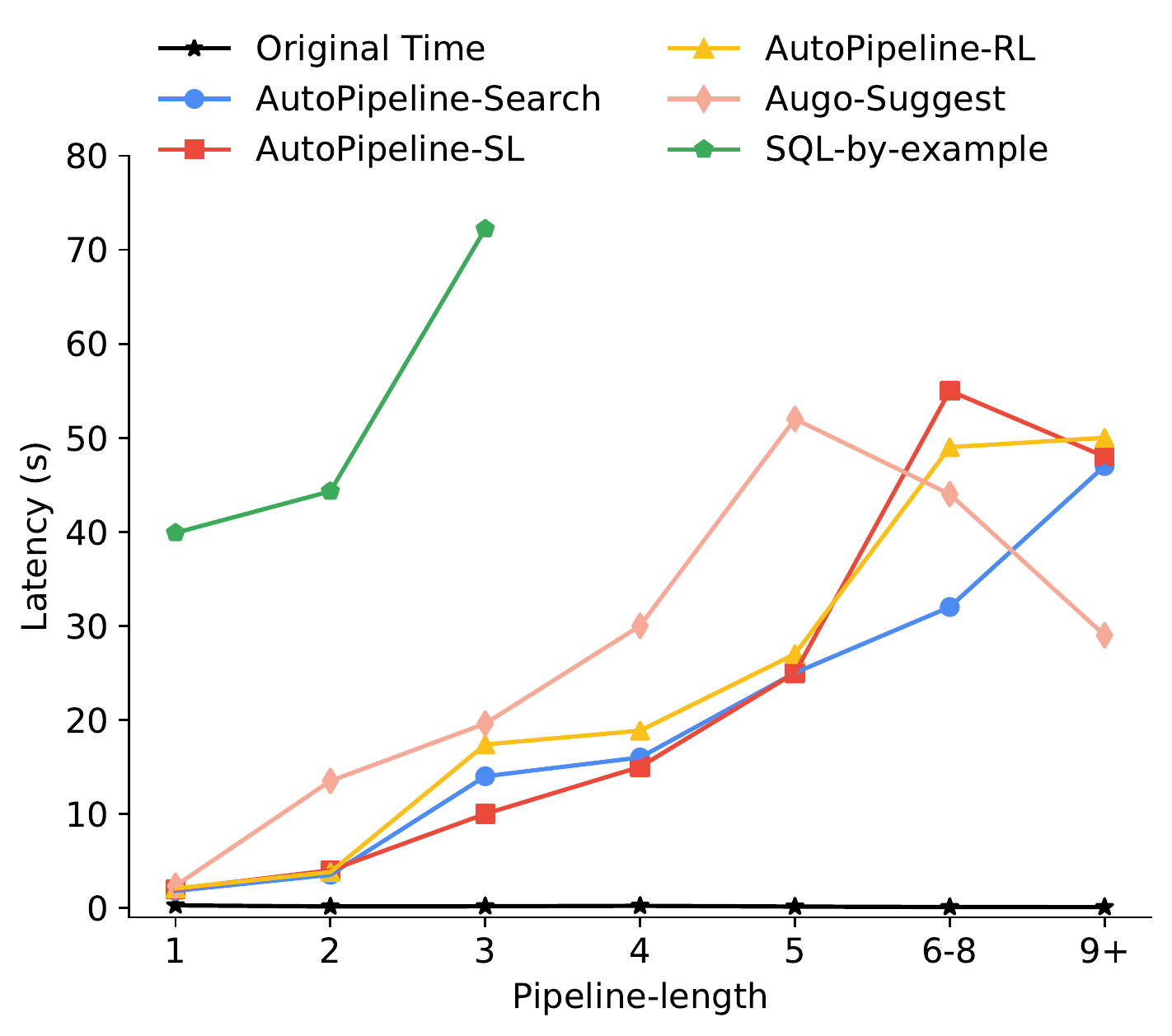}
  \vspace{-3mm}
  \caption{Latency results by pipeline-lengths on the \textsf{GitHub} benchmark}
    \label{fig:latency-comp}
\endminipage
\vspace{-3mm}
\end{figure*}

\textbf{Synthesis Quality.}
Figure~\ref{fig:acc-comp} and Figure~\ref{fig:mrr-comp}
show detailed comparisons of
accuracy and MRR, between different
methods on the \textsf{GitHub} benchmark.
Test pipelines are bucketized into 7 groups based on their
lengths (shown on the x-axis), with longer pipelines being
more difficult to synthesize. It can be seen
that \sjrl{} and \sjs{} are comparable in terms of quality,
with \sjrl{} being slightly better in terms of MRR. We can see
that \sjrl{} is noticeably better than \sjsl{}, showing the benefit of
using RL to proactively select
examples to learn from.

All \sj{} variants are substantially better
than \textsf{QBO} and \textsf{SQL-by-example} baselines.
We note that \textsf{SQL-by-example} fails to synthesize
any pipeline longer than 3-steps within our 1 hour timeout limit,
showing its limited efficacy when dealing with
large input tables from real pipelines.

\textbf{Latency.}
Figure~\ref{fig:latency-comp} shows a comparison of the
average latency to successfully synthesize a pipeline between all methods
on the \textsf{GitHub} benchmark. While all \sj{} methods have
comparable latency, \textsf{SQL-by-example}
requires 20x-7x more time to synthesize pipelines up to 3 steps.

\begin{figure}[h]
        \centering
        \subfigure[A human-authored pipeline]{
            \includegraphics[width=\columnwidth]{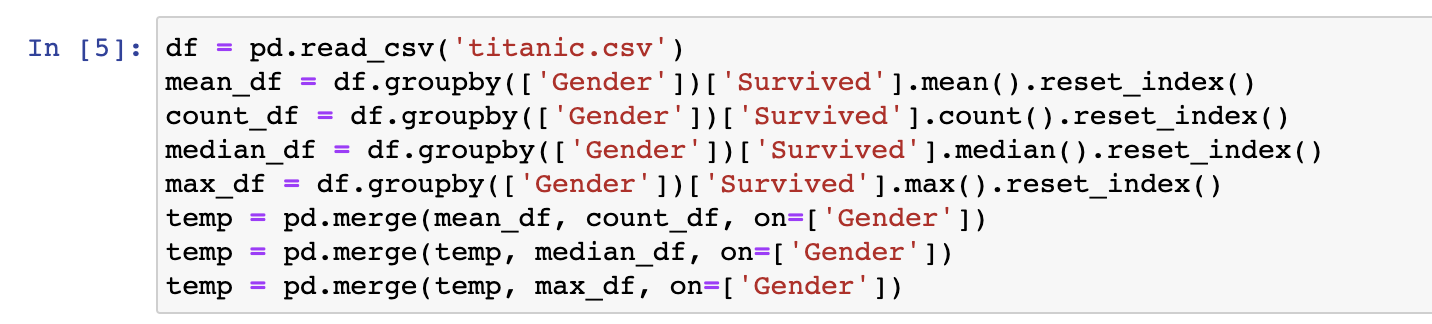}
            \label{fig:beforeSim}
            }\vspace{-3mm}
        \subfigure[A synthesized pipeline with fewer steps]{
            \includegraphics[width=\columnwidth]{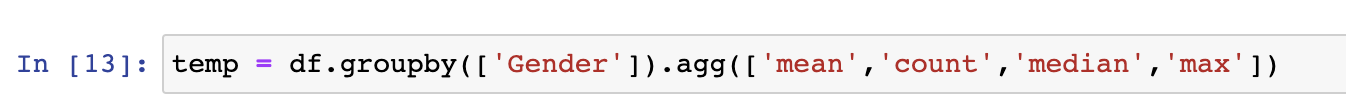}
            \label{fig:afterSim}
            }
            \vspace{-3mm}
        \caption{An example pipeline simplified after synthesis.}
\vspace{-3mm}
        \label{fig:ppl-sim}
    \end{figure}

\textbf{Pipeline simplification.} We observe in experiments that
our synthesized pipelines can sometimes be simpler (with fewer
steps) than human-authored ground-truth
pipelines, while still being
semantically equivalent. Figure~\ref{fig:ppl-sim}
shows a real example from GitHub, where
the human-authored pipeline would
group-by on column \texttt{Gender} for four times with
different aggregation, 
before joining them back.
A synthesized pipeline from \sj{} is a one-liner in this
case and more succinct. While this example
is intuitive, there are many more involved examples of
simplifications -- for example, having an Unpivot on
each of K similar files followed by (K-1) union, is equivalent to
a Join between the K similar files
followed by one Unpivot, etc.

Out of the 700 pipeline in the
\textsf{GitHub} benchmark, our synthesized pipelines are
simpler on 90 cases (12.85\%), which we
believe is an interesting use and
an added benefit of \sj. 

  \begin{figure*}
\vspace{-15mm}
    \minipage{0.3\textwidth}
      \includegraphics[width=\linewidth]{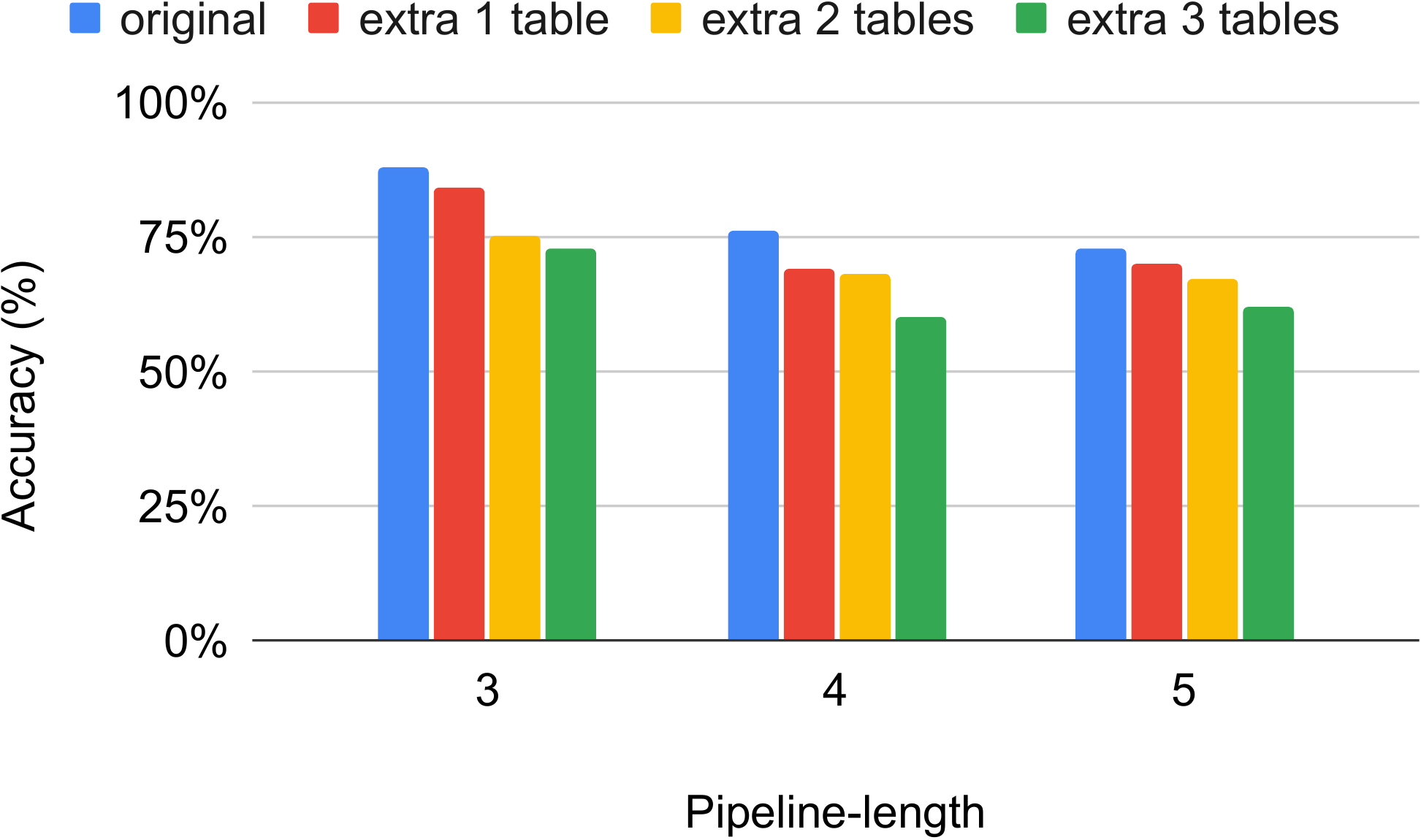}
        \caption{Robustness: add extra input tables irrelevant to pipelines.}
        \label{fig:table-robust}
    \endminipage\hfill
    \minipage{0.3\textwidth}
      \includegraphics[width=\linewidth]{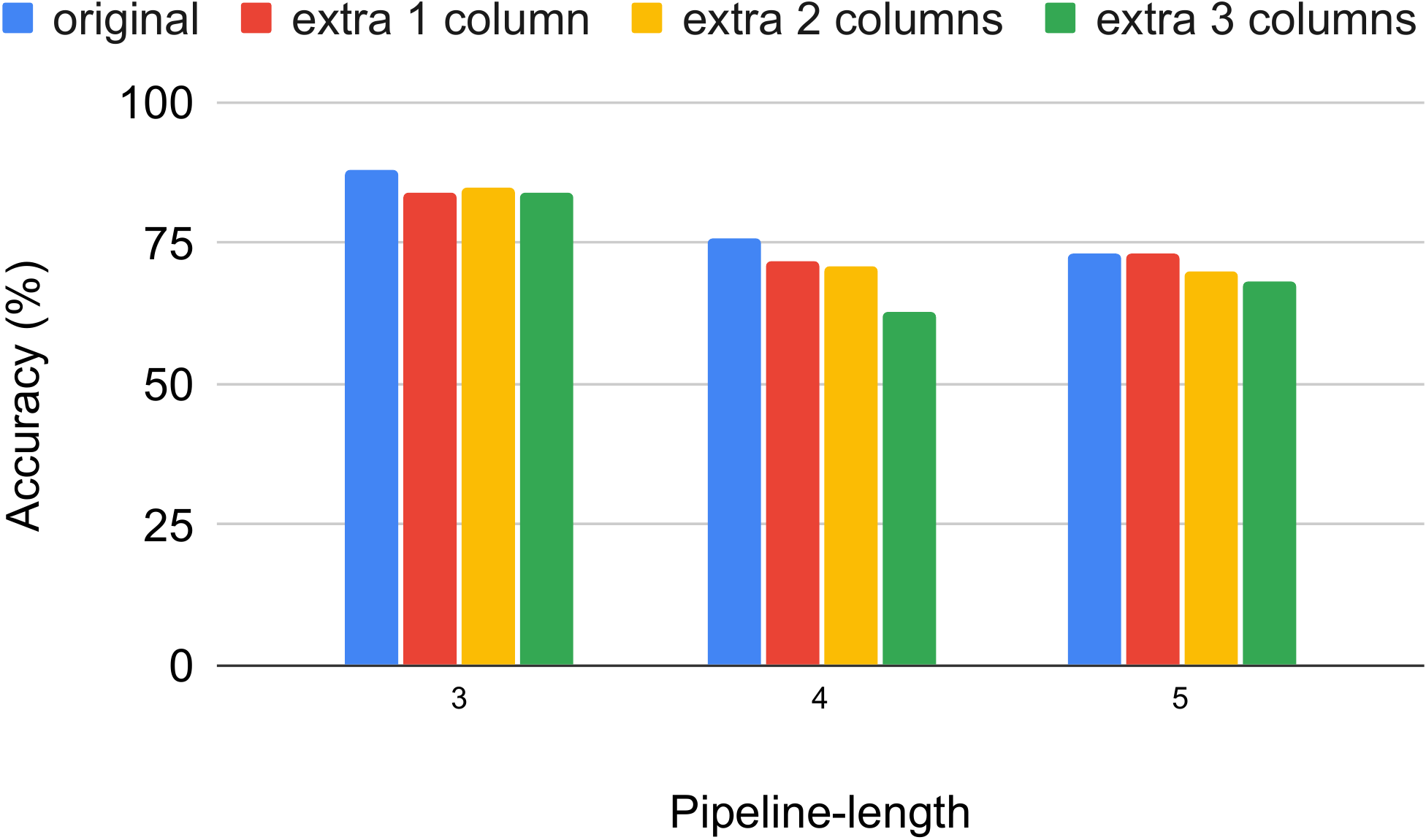}
        \caption{Robustness: add extra columns irrelevant to pipelines.}
      \label{fig:column-robust}
    \endminipage\hfill
    \minipage{0.3\textwidth}%
      \includegraphics[width=\linewidth]{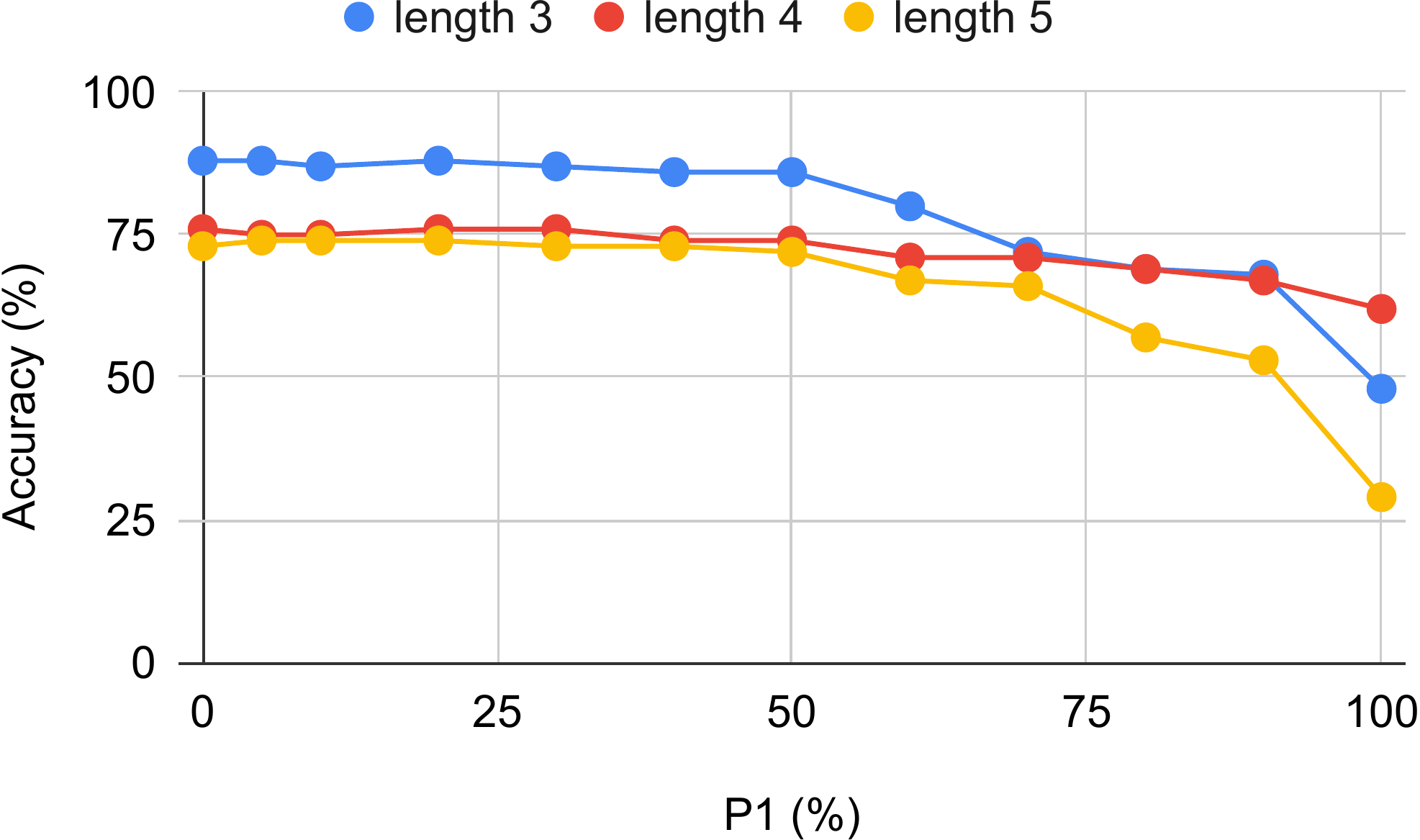}
        \caption{Robustness: randomly perturb column values.}
      \label{fig:perturb-robust}
    \endminipage
    \end{figure*}

\subsection{Robustness Analysis}
\label{sec:robustness}
To understand the robustness of our algorithm in the
presence of noisy input/output tables,
we perform various robustness tests.

\textbf{Add irrelevant input tables.}
For each pipeline synthesis task, we inject $K$ extra
input tables randomly sampled from other test pipelines
irrelevant to this synthesis task, which is used to
test the robustness of our algorithm in the presence
of irrelevant data sources.
Figure~\ref{fig:table-robust} shows that even with
3 extra irrelevant input tables, our synthesis algorithm
(not specifically optimized to handle noisy and irrelevant tables)
only has a slight decrease in accuracy.

\textbf{Add irrelevant columns to input tables.} In addition
to injecting extra irrelevant tables, for each test pipeline,
we also test the scenario
in which we inject $K$ extra columns to each input
table randomly sampled from
other irrelevant pipelines.
Figure~\ref{fig:column-robust} shows that our
accuracy also drops slightly in such settings.

\textbf{Perturb column values.} In order to test the robustness
of our synthesis in the presence of noisy data values (e.g.,
typos and name variations), for each input table in a
pipeline, we randomly select 50\% of
string columns and randomly perturb $p$ fraction of their values.
Specifically, we randomly initialize a character scrambling scheme (e.g.,
`a' $\rightarrow$ `t', `b' $\rightarrow$ `f', etc.), and for $p$
fraction of distinct values in a selected column, we apply the scrambling
character-by-character to perturb values
(e.g., `abc store' $\rightarrow$ `tfg kabzo'). Such perturbations
are performed on each input table independently.
Figure~\ref{fig:perturb-robust}
shows the synthesis accuracy when varying $p_1$ from 0 to 100\%.
It can be seen that our synthesis is still robust
even when 50\% of values are perturbed.


\begin{figure}
  \vspace{-5mm}
\minipage{0.23\textwidth}
  \includegraphics[width=\linewidth]{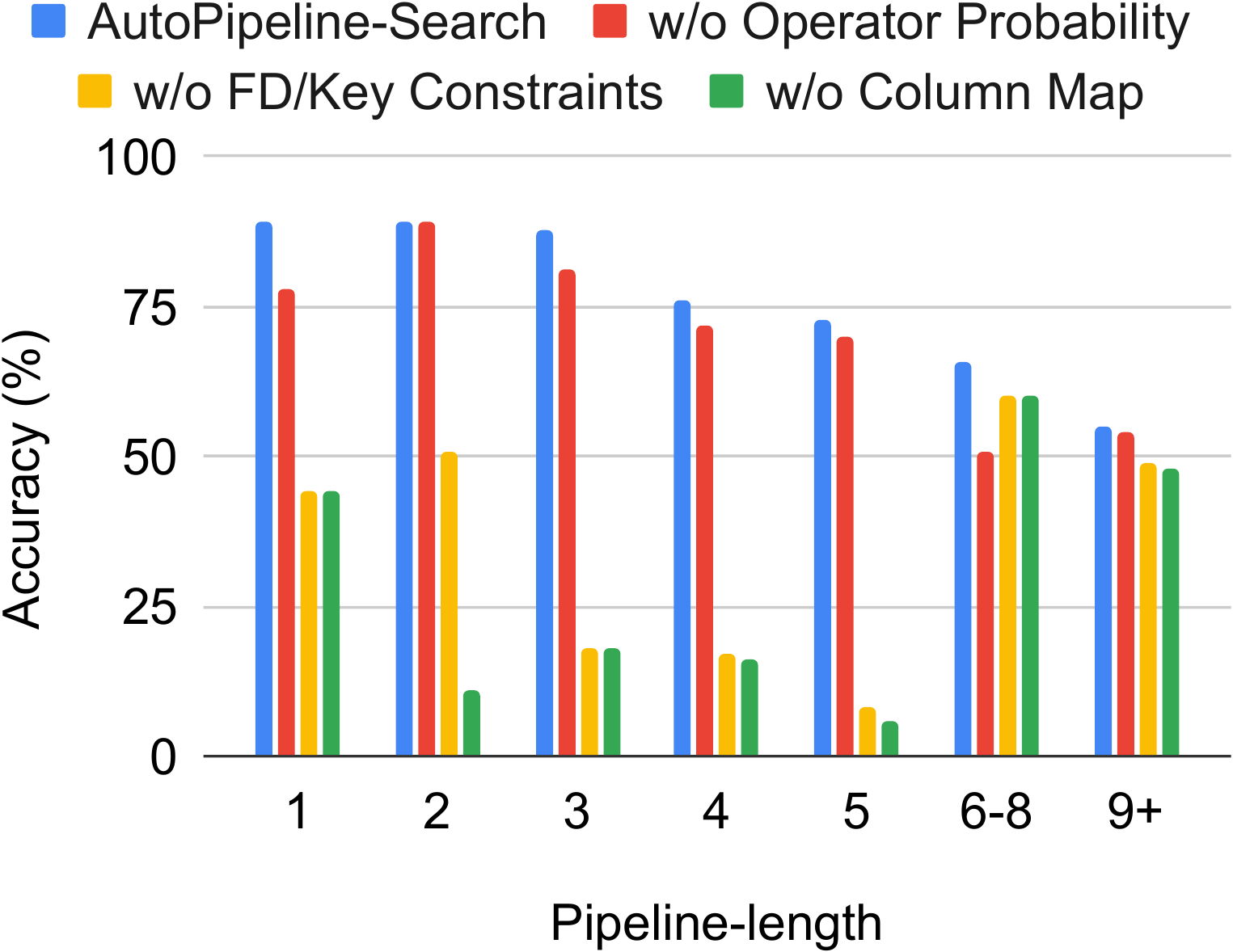}
  \vspace{-5mm}
  \caption{Ablation study.}
	\label{fig:ablation-accuracy}
\endminipage\hfill
\minipage{0.23\textwidth}
  \includegraphics[width=\linewidth]{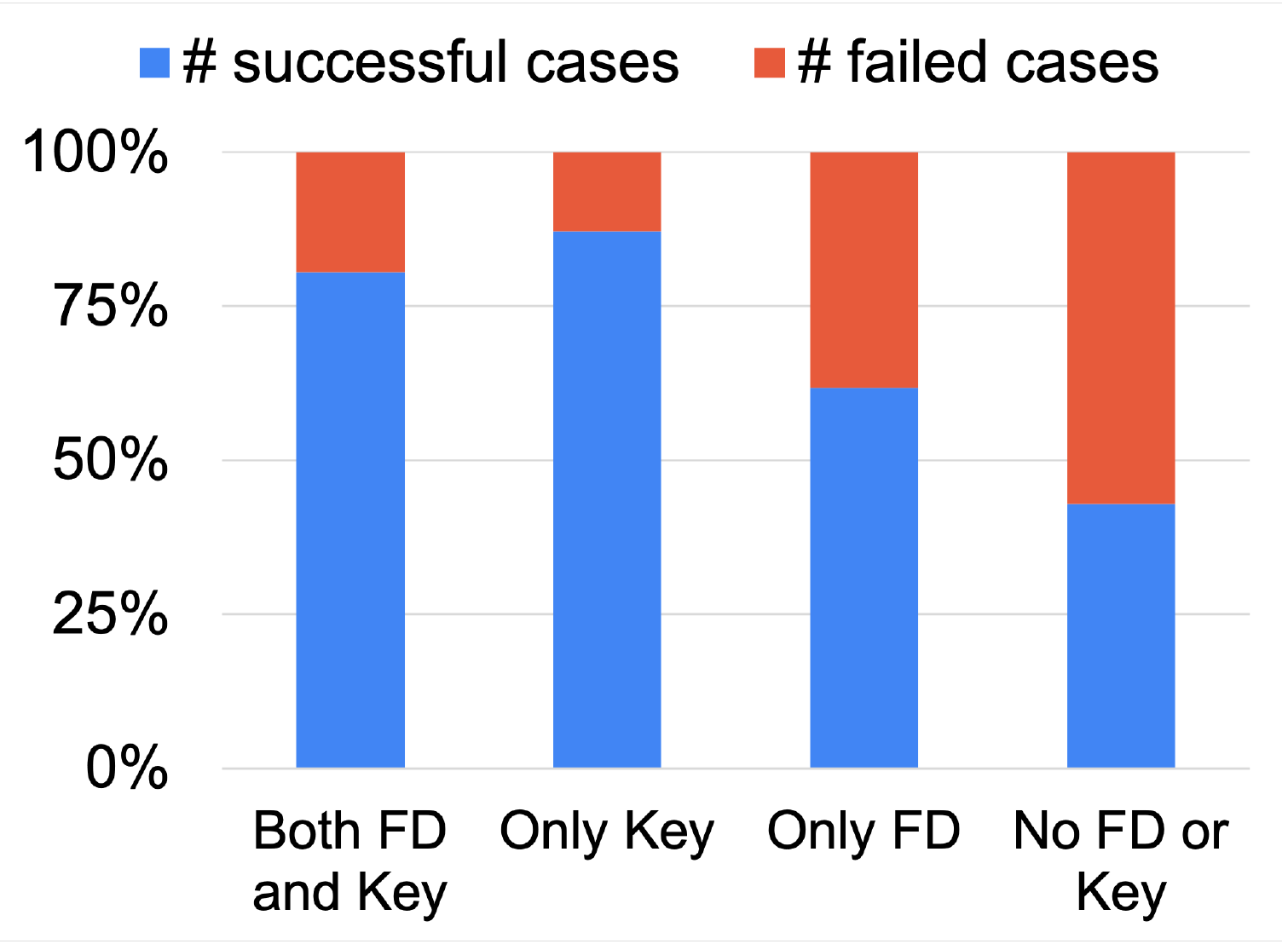}
\vspace{-5mm}
    \caption{Error Analysis.}
	\label{fig:error-analysis}
\endminipage
\vspace{-3mm}
\end{figure}

\vspace{-2mm}
\subsection{Error Analysis}
\label{sec:error}
In this section, we analyze 70 sampled failed cases
(10 cases for each group of pipeline lengths), to
understand why \sj{} fails to synthesize.
We categorize the failed reasons as follows:
\begin{itemize}[leftmargin=*]
\item \underline{Incorrect singe operator parameter}. 41\% of failed cases
fall in this category (e.g., a ground-truth join column is not in the top-K
parameters predicted for the given pair of tables).

\item \underline{Incorrect string transformation}. 26\% failed
cases are in this category, which can be
attributed to the fact that the synthesis of string
transformations in our case do not have paired input/output examples
(unlike the traditional by-example setting).

\item \underline{False-positive FDs}.
There are two cases where false-positive FDs are discovered
from target tables, which prevents synthesized pipelines to
produce matching FDs, causing the synthesis to fail.

\item \underline{Deleted key column.}  In three test cases, key columns are
deleted in final steps of the pipelines, leading to missing constraints.
\end{itemize}

\textbf{Spurious discovery of constraints.}
While it is known that spurious FDs/keys
can be discovered~\cite{papenbrock2016hybrid},
especially on small input tables,
they do not contribute significantly to failed synthesis
in our evaluation. We believe
this is because unlike by-example synthesis that
uses only a few rows, our test pipelines operate on
real data tables (e.g., from Kaggle) that are typically large
(e.g., on average an input table used in
GitHub pipelines have over 4K rows, and an output
table has over 41K rows). Such large tables make
false discovery of spurious constraints substantially less likely.
To confirm this, we down-sample target tables
of each test case to 20 rows each and re-run \sj.
We observe 37 out of 70
cases would then fail due to spurious FDs/Keys,
confirming the hypothesis that large realistic
tables indeed prevents spurious discovery of constraints.

\textbf{Contribution of constraints in synthesis.}
Figure~\ref{fig:error-analysis} shows how
fail-rates vary when FDs/Keys do not
exist in the target table $T^{tgt}$. We can see from the
two rightmost bars that
when only FDs exist $T^{tgt}$
(key columns are missing),
or when both FDs/Keys are missing, fail-rates go up significantly.

\iftoggle{fullversion}{
}
{
\textbf{Additional results.} We present
additional results such as sensitivity analyses of
in a full version of the paper~\cite{full}.

\begin{figure}
  \vspace{-1mm}
  \centering
  \includegraphics[width=1\columnwidth]{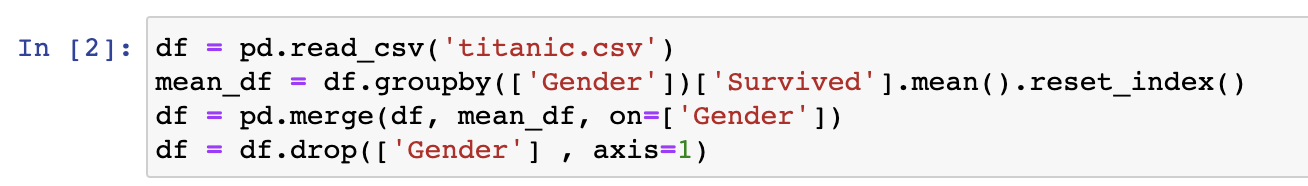}
  \vspace{-5mm}
  \caption{An example pipeline that we fail to synthesize.}
  \label{fig:error-pipeline}
  \vspace{-5mm}
\end{figure}
}

%% file: related.tex

\section{Related Works}
\label{sec:related}

\iftoggle{fullversion}{We review related work in the broad
area of automating pipeline-building and data preparation.}
{
We briefly review related work on 
automating pipeline-building in this section in the interest
of space. 
We give additional discussion of the broad
area of data preparation in a full version of the paper~\cite{full}.
}

\textbf{Automate data transformations.}
Data transformation is a long-standing problem and a common
step in data-pipelines. Significant progress has been made
in this area, with recent work for ``by-pattern''~\cite{jin2020auto}
and ``by-example'' 
approaches~\cite{Abedjan16, gulwani2011, He2018TDE, he2018transform, Heer15, Jin17, jin2018clx, Singh16}.
These techniques have generated substantial impacts on 
commercial systems, with related features shipping 
in popular systems such as
Microsoft Excel~\cite{gulwani2011}, Power BI~\cite{He2018TDE}, and
Trifacta~\cite{TBE-trifacta}.

\iftoggle{fullversion}{
\textbf{Automate table joins.}
Join is an important operator that
brings multiple tables together. However, tables in the wild
often use different value representations, making
equi-join insufficient. Various statistical
and program-synthesis-based methods have been 
proposed to infer pairs of values to join that are
not exact equi-joins~\cite{search-join, auto-join, he2015sema, auto-fj, Warren06, HassanzadehPYMPHH13}. 

More broadly, entity-matching 
(EM)~\cite{christophides2019end, getoor2012entity}
can be considered as generalized forms of table joins
and thus relevant in pipeline-building,
where all table columns across two tables 
may contribute to match/non-match
decisions (as opposed to only selected join columns
in the case of join). A large number of methods
have been developed to automate 
EM, including recent approaches
such as~\cite{mudgal2018deep, li2020deep, 
zhao2019auto, zhang2020multi, li2021deep, gurajada2019learning}.

\textbf{Automate table restructuring.} Table restructuring
operations (e.g., Pivot, Unpivot, Transpose, etc.)
are also challenging steps in pipeline-building, where
by-example approaches~\cite{barowy2015flashrelate, Jin17}
have been developed in the research literature.
Because it is difficult for users to specify an entire 
output table by-example, recommendation-based 
approaches~\cite{auto-suggest} have also been proposed,
in which top-K likely table-restructuring options are recommended
without requiring users to provide example output tables.

\textbf{Automate error detection and repair.}
Detecting data errors and fixing them is also an important
part of the overall data preparation process, which can
also be automated to various extents leveraging
user-provided constraints~\cite{yan2020scoded, chu2013holistic, rekatsinas2017holoclean, fan2014interaction, cong2007, hellerstein2008}, or
a small number of labeled examples~\cite{mahdavi2020baran, mahdavi2021semi, lahijani2020semi, heidari2019holodetect}.
Unsupervised methods~\cite{wang2019uni, 
huang2018auto} have also been developed
in the context of Excel-like spreadsheets, at the cost of lower 
coverage/recall.

Because error detection and repair methods tend to generate
\textit{local} changes (e.g., applicable to some selected 
cells) and typically require
humans to verify, we consider these steps as orthogonal to
the end-to-end pipeline automation problem we 
study, where changes are typically
more \textit{global} (e.g., applicable to all cells in the same column),
and do not require human users to verify case-by-case/cell-by-cell.
}
{
}

\textbf{Multi-step pipeline synthesis.} 
As discussed, existing work on multi-step pipelines
mostly focus on the by-example paradigm
(e.g., SQL-by-example~\cite{sql-by-example},
Query-by-output~\cite{qbo}, and 
Auto-Pandas~\cite{bavishi2019autopandas}), 
where an exact output-table is often difficult for users to provide.
In addition, most approaches
support a limited set of operators, which limits
their ability to synthesize complex pipelines.

\iftoggle{fullversion}{
\textbf{Code search.} 
Code search (e.g., \cite{mukherjee2020searching, reiss2009semantics}) 
is another line
of research relevant to pipeline synthesis. While we also synthesize
code in the context of data pipelines, our synthesis is mostly guided
by input/output data tables, which is thus orthogonal to the
code context leveraged by code search systems.
}
{
}

%% file: conclusion.tex
\section{Conclusions and Future Works} \label{sec:conclusion}
In this paper, we propose a new by-target synthesis paradigm
to automate pipeline-building.
We design search and learning-based synthesis
algorithms, which are shown to be effective on real data pipelines.
Future directions include extending our current DSL,
and incorporating user feedback to facilitate synthesis.